\documentclass[twocolumn]{aastex63}

\usepackage{graphicx}
\usepackage{amsmath}
\usepackage{breqn}
\usepackage{amssymb}
\usepackage{hyperref}
\usepackage{xcolor}
\usepackage{comment}

\usepackage{graphics}

\usepackage{amsmath}
\usepackage{amssymb}

\usepackage{mathrsfs}

\newcommand{\seedmass}{M_{\mathrm{seed}}}

\newcommand{\seedmassdynamical}{M^{\mathrm{dyn}}_{\mathrm{seed}}}

\newcommand{\mh}{\tilde{M}_{\mathrm{h}}}
\newcommand{\msfmp}{\tilde{M}_{\mathrm{sfmp}}}

\submitjournal{ApJ}

\begin{document}
\title[Dynamics of low-mass black hole seeds]{Dynamics of low-mass black hole seeds in the \texttt{BRAHMA} simulations using subgrid-dynamical friction: Impact on merger-driven black hole growth in the high redshift Universe }

\correspondingauthor{Aklant K. Bhowmick}
\email{aklant.app@gmail.com}

\author[0000-0002-7080-2864]{Aklant K. Bhowmick}
\affiliation{Department of Astronomy, University of Virginia, Charlottesville, VA 22904, USA} 

\author[0000-0002-2183-1087]{Laura Blecha}
\affiliation{Department of Physics, University of Florida, Gainesville, FL 32611, USA}

\author{Luke Z. Kelley}
\affiliation{Department of Astronomy, University of California Berkeley, Gainesville, FL 32611, USA}

\author{Aneesh Sivasankaran}
\affiliation{Department of Physics, University of Florida, Gainesville, FL 32611, USA}

\author[0000-0002-5653-0786]{Paul Torrey}
\affiliation{Department of Astronomy, University of Virginia, Charlottesville, VA 22904, USA}
\affiliation{Virginia Institute for Theoretical Astronomy, University of Virginia, Charlottesville, VA 22904, USA}
\affiliation{The NSF-Simons AI Institute for Cosmic Origins, USA}

\author[0000-0001-6260-9709]{Rainer Weinberger}
\affiliation{Leibniz Institute for Astrophysics Potsdam (AIP), An der Sternwarte 16, 14482 Potsdam, Germany}

\author{Nianyi Chen}
\affiliation{School of Natural Sciences, Institute for Advanced Study, Princeton, NJ 08540, USA}

\author[0000-0001-8593-7692]{Mark Vogelsberger}
\affiliation{Department of Physics and Kavli Institute for Astrophysics and Space Research, Massachusetts Institute of Technology, Cambridge, MA 02139, USA}

\author[0000-0001-6950-1629]{Lars Hernquist}
\affiliation{Harvard-Smithsonian Center for Astrophysics, Harvard University, Cambridge, MA 02138, USA}

\author[0000-0001-6950-1629]{Priyamvada Natarajan}
\affiliation{Dept. of Astronomy, Yale University, 219 Prospect Street, New Haven, CT 06511, USA}
\affiliation{Dept. of Physics, 217 Prospect Street, Yale University, New Haven, CT 06511,USA}
\affiliation{Black Hole Initiative, Harvard University, 20 Garden Street, Cambridge, MA 02138, USA}

\begin{abstract}
We analyze the dynamics of low-mass black hole (BH) seeds in the high-redshift~($z\gtrsim5$) Universe using a suite of $[4.5~\mathrm{Mpc}]^3$ and $[9~\mathrm{Mpc}]^3$  \texttt{BRAHMA} cosmological hydrodynamic simulations. The simulations form seeds with mass $\seedmass=2.2\times10^3~M_{\odot}$ in halos that exceed critical thresholds of dense~\&~metal-poor gas mass~($5-150~ \seedmass$) and the halo mass~($1000-10000~\seedmass$). While the initial \texttt{BRAHMA} boxes pinned the BHs to the halo centers, here we implement a sub-grid dynamical friction (DF) model. We also compare simulations where the BH is allowed to wander without the added DF. We investigate the spatial and velocity offsets of BHs in their host subhalos, as well as BH merger rates. We find that subgrid DF is crucial to ensure that a significant fraction of BHs effectively sink to halo centers by $z\sim5$, thereby enabling them to get gravitationally bound and merge with other BHs at separations close to the spatial resolution~($\sim0.2-0.4~\rm kpc$) of the simulation. For the BHs that merge, the associated merger time scales lag between $\sim100-1000~\mathrm{Myr}$ after their host halos merge. Compared to predictions using BH repositioning, the overall $z\gtrsim5$ BH merger rates under subgrid DF decrease by a factor of $\sim4-10$.
Under subgrid DF, the different seed models predict merger rates between $\sim100-1000$ events per year at $z\gtrsim5$. These mergers dominate early BH growth, assembling BHs up to $\sim10^4-10^5~M_{\odot}$ by $z\sim5$, wherein $\lesssim2~\%$ of their mass is assembled via gas accretion. Our results highlight the promise for constraining seeding mechanisms using gravitational waves from future facilities such as the Laser Interferometer Space Antenna. 
 \end{abstract}


\section{Introduction}
The origin of supermassive black holes~(SMBHs) with masses $\gtrsim 10^6~M_{\odot}$ 
continues to be a challenging open question at present. Proposed theoretical models trace their origin to the formation of smaller ``seed" black holes  that are born out of standard astrophysical processes in the very early Universe. These seeds typically fall into the following three broad categories: ``light"~($\sim10-10^3~M_{\odot}$) seeds as Population III stellar remnants or Pop III seeds~\citep{2001ApJ...550..372F,2001ApJ...551L..27M,2013ApJ...773...83X,2018MNRAS.480.3762S}, ``heavy" seeds~($\sim10^4-10^5~M_{\odot}$) as Direct Collapse Black Holes (DCBH) seeds~\citep{2003ApJ...596...34B,2006MNRAS.370..289B,2006MNRAS.371.1813L,2007MNRAS.377L..64L,2018MNRAS.476.3523L,2019Natur.566...85W,2020MNRAS.492.4917L,2023MNRAS.526L..94B}, and ``medium-weight" seeds~($\sim10^3-10^4~M_{\odot}$) as remnants of runaway stellar collisions in dense nuclear star clusters (NSC) seeds~\citep[][]{2011ApJ...740L..42D,2014MNRAS.442.3616L,2020MNRAS.498.5652K,2021MNRAS.503.1051D,2021MNRAS.tmp.1381D,2021MNRAS.501.1413N}. Medium-weight seeds could also form via accelerated accretion onto light seeds in gas-rich NSCs~\citep{2014Sci...345.1330A}, processes that have been shown to occur throughout cosmic time \citep{2021MNRAS.501.1413N}. Beyond these standard ``astrophysical" seeds, other types of seeds have also been suggested to form out of more ``exotic" processes such as dark matter annhilation~\citep{2019MNRAS.483.3592B,2023MNRAS.525..969S,2024arXiv241201828T,2025MNRAS.536..851C}, dark-matter self interactions~\citep{2019JCAP...07..036C,2023PhRvD.107h3010M,2025JCAP...01..060G,2025arXiv250400075S,2025arXiv250323710J} and from primordial black holes~\citep{2001JETP...92..921R,2005APh....23..265K, 2022ApJ...926..205C,2024arXiv241103448Z,2024SCPMA..6709512Y}.

The seed masses in most of the above models lie in the ``intermediate mass black hole"~(IMBH) regime of $\sim10^2-10^5~M_{\odot}$. However, direct detection of IMBHs using electromagnetic~(EM) observatories is extremely challenging, particularly at high redshifts. 
Even with powerful facilities like the James Webb Space Telescope~\citep[JWST,][]{2023PASP..135e8001M} and future X-ray facilities like the Advanced X-ray Imaging Satellite~\citep[AXIS,][]{2023SPIE12678E..1ER}, BHs down to $\sim10^5~M_{\odot}$ will be hard to detect at high redshifts. This leaves the bulk of the IMBH populations at early cosmic epochs virtually inaccessible to EM observations. To that end, gravitational wave~(GW) observations have now emerged as one of the most promising probes for detecting IMBHs as they merge. The Laser Interferometer Gravitational Wave Observatory~\citep[LIGO][]{2009RPPh...72g6901A} recently made the first detection of an IMBH in GW190521, which was a merger of two BHs that produced a  $\sim142~M_{\odot}$ remnant~\citep{2020ApJ...900L..13A}. 

Although LIGO is only sensitive to merging BHs up to a few $\sim100~M_{\odot}$ in the nearby Universe, the upcoming Laser Interferometer Space Antenna~(LISA) is expected to detect the bulk of the IMBH and SMBH populations ranging in mass from 
$\sim10^3~M_{\odot}$ to $\sim10^7~M_{\odot}$~\citep{2009RPPh...72g6901A} out to $z\sim15$~\citep{2017arXiv170200786A}. A growing body of literature suggests that the GW event rates detected by LISA are expected to be strongly impacted by  seeding~\citep{2007MNRAS.377.1711S,2012MNRAS.423.2533B,2018MNRAS.481.3278R,2019MNRAS.486.2336D,2020MNRAS.491.4973D,2021MNRAS.507.2012B,2024MNRAS.531.4311B}. Furthermore, \cite{2018MNRAS.481.3278R} demonstrated for the first time that LISA is critical for disentangling signatures of seeding from accretion and dynamics. 


Robust predictions for LISA gravitational wave event rates require comprehensive models of massive black hole formation, dynamics, and mergers within a full cosmological context. Cosmological simulations have become a powerful framework for developing such models, especially given their recent success in reproducing a wide range of observed galaxy properties~\citep{2012ApJ...745L..29D,2014Natur.509..177V,2015MNRAS.452..575S,2015MNRAS.450.1349K,2015MNRAS.446..521S,2016MNRAS.460.2979V,2016MNRAS.463.3948D,2017MNRAS.467.4739K,2017MNRAS.470.1121T,2019ComAC...6....2N,2020MNRAS.498.2219V,2020NatRP...2...42V,2024arXiv240910666N}. However, modeling and tracking \textit{both} the formation and the dynamics of BHs has proven to be extremely challenging in cosmological hydrodynamic simulations due to resolution limitations. 

In terms of modeling ab-initio seed formation, the initial mass of a BH seed that can form in the box is limited by the gas mass resolution and the ability to track gravitational collapse.  The gas mass resolutions in most of the larger volume simulations~(with $\sim25~\mathrm{Mpc}$ to a few Gpc box-lengths) are typically $\sim10^5-10^7~M_{\odot}$ -- larger than the seed masses predicted by almost all postulated seeding channels. Therefore, many simulations leverage the general observational expectation that massive galaxies harbor SMBHs to simply start with fairly massive seeds~($\sim10^5-10^6~M_{\odot}$) that are simply assigned by birth to halos based on a halo mass threshold criterion~($\sim10^9-10^{10}~M_{\odot}$)
~\citep[for e.g.][]{2014Natur.509..177V,2015MNRAS.450.1349K,2016MNRAS.455.2778F,2019ComAC...6....2N}.
Although such simulations are agnostic to the ab-initio formation of seeds and their early growth, they have proven to be useful for LISA event rate predictions, as they can track the merger of massive $\sim10^5-10^7~M_{\odot}$ BHs. However, they do not capture a large proportion of expected LISA populations comprising $\sim10^3-10^5~M_{\odot}$ seed BHs and therefore cannot discriminate between the different seed formation channels. Several more recent simulations have meanwhile adopted more sophisticated seed models that are based on local gas properties~\citep{2015MNRAS.448.1835T,2017MNRAS.470.1121T,2017MNRAS.468.3935H,2017MNRAS.467.4739K}. 
While the new criteria present a significant improvement over the simple halo-mass based seeding prescriptions, it is critical to perform systematic studies of how different seed models impact the merging BH populations to make robust predictions for LISA.


To this end, we recently developed the \texttt{BRAHMA} cosmological simulation suite which forms the basis for the largest systematic study to quantify the impact of seeding on BH populations across cosmic time~\citep{2024MNRAS.531.4311B,2024MNRAS.533.1907B,2024arXiv241119332B}. Before \texttt{BRAHMA}, these systematic studies had been carried out mainly using semi-analytic models~\citep[SAMs][]{2012MNRAS.423.2533B,2018MNRAS.481.3278R,2019MNRAS.486.2336D,2021MNRAS.506..613S,2022MNRAS.511..616T,2023MNRAS.519.4753T,2023ApJ...946...51C,2023MNRAS.518.4672S}. The \texttt{BRAHMA} simulations implement a wide range of parametrized seeding model variants that depend on gas properties identified to be important for the Pop III, NSC and DCBH seed formation channels, such as high gas density, low gas metallicity, strong Lyman-Werner radiation, low gas spins, and rich environments. These seeding prescriptions have been extensively tested for resolution convergence and the ability to reproduce local BHs and high-z quasar populations using a large suite of zoom and constrained simulations~\citep{2021MNRAS.507.2012B,2022MNRAS.510..177B,2022MNRAS.516..138B}. 

In addition to a seed model, a robust BH dynamical model for merging is crucial for tracking the evolution of $\sim10~\rm kpc$ separation BH pairs produced by galaxy mergers to separations of $\lesssim 0.001~\rm pc$ whence loss of orbital angular momentum via GW emission drives BH coalescence. This journey to sub-pc scales is thought to be driven by several physical  processes that operate on distinct physical scales: BH dynamical friction~(DF) from $\sim10 - 0.1~\rm kpc$ separations; stellar loss cone scattering~\citep{Merritt_2013} from $\rm \sim0.1~kpc-1~pc$ separations; viscous drag from gas in circumbinary disks~\citep{2002ApJ...567L...9A,2023MNRAS.522.2707S} from $\rm \sim1~pc - 0.001~pc$ separations, and finally GW emission at $\lesssim 0.001~\rm pc$ separations~\citep[][see Figure 2]{2021MNRAS.501.2531S}. 

However, in cosmological simulations, even the earliest stages of BH inspiral cannot be effectively resolved because of the inability to capture the BH DF. This is due to 
the coarseness of the background field of gas, stars and DM particles, as well as the softening of the gravitational force at small scales that suppresses the two-body interactions responsible for driving DF. Capturing DF becomes extremely problematic when BH masses are close to or smaller than the background particle masses, leading to spuriously large two-body interactions that can ``numerically heat" the BHs and artificially enhance their wandering~\citep{2022MNRAS.516..167B}. The vast majority of the simulations described above deal with this issue by simply ``pinning" or ``repositioning" the BHs to the local potential minima in their host halos. This is precisely what has been done so far in the \texttt{BRAHMA} simulations. However, this BH repositioning scheme inevitably leads to unrealistically prompt mergers of BHs, or spurious mergers during fly-by encounters amongst halos. Therefore, the resulting merger rates from the original \texttt{BRAHMA} simulations~\citep{2024MNRAS.531.4311B,2024MNRAS.533.1907B} can at best be interpreted as the upper limits of the merger rates for a given seed model. 

In recent years, several prescriptions have been developed to correct for the missing dynamical friction force~\citep{2015MNRAS.451.1868T,2019MNRAS.486..101P,2022MNRAS.510..531C,2023MNRAS.519.5543M,2024A&A...692A..81D,2024MNRAS.534..957G} in cosmological simulations. 
Unlike the repositioning methods, they 
capture the inspiral of BH pairs after their galaxies have already merged. Moreover, they can capture the possibility of BHs wandering away from galaxy centers~\citep{2021MNRAS.503.6098R,2021MNRAS.505.5129B,2023MNRAS.525.1479D}, which has been confirmed with observations of dwarf galaxies~\citep{2020ApJ...888...36R}.  

In this work, we combine our seed models with a subgrid DF model to study the dynamics and merger time-scales of low-mass $\sim10^3~M_{\odot}$ seeds forming in low mass halos~($\sim10^6-10^7~M_{\odot}$) at high-redshifts~($z\gtrsim5$). The pairing of this subgrid DF model with the \texttt{BRAHMA} seeding framework facilitates increasingly accurate BH merger rate predictions for LISA. Here we start by applying the \citealt{2023MNRAS.519.5543M}~(hereafter M23) model. 
The application of this model allows us to trace the evolution of merging BH pairs all the way down to the spatial resolution of our simulations of $\sim0.2-0.4~\rm kpc$; i.e. 2 times the gravitational softening length. This enables us to study the formation rates of $\sim0.2-0.4~\rm kpc$ scale BH pairs, and compare against our prior results using BH repositioning~\citep{2024MNRAS.533.1907B} which could only probe BH pairs at significantly larger scales~($\sim1-15~\rm kpc$). 

The structure of this paper is as follows.  Section \ref{methods} discusses the simulation suite and models for BH seeding and dynamics. Section \ref{results} describes the results, starting with the dynamics of the overall population of BHs and followed by the dynamics of merging BH pairs. In Section \ref{discussion}, we discuss the implications of our results in the context of similar works in the literature, and we summarize the main conclusions in Section \ref{Conclusions}.

\section{Methods}
\label{methods}
\begin{figure*}
\centering
\includegraphics[width=0.85\textwidth]{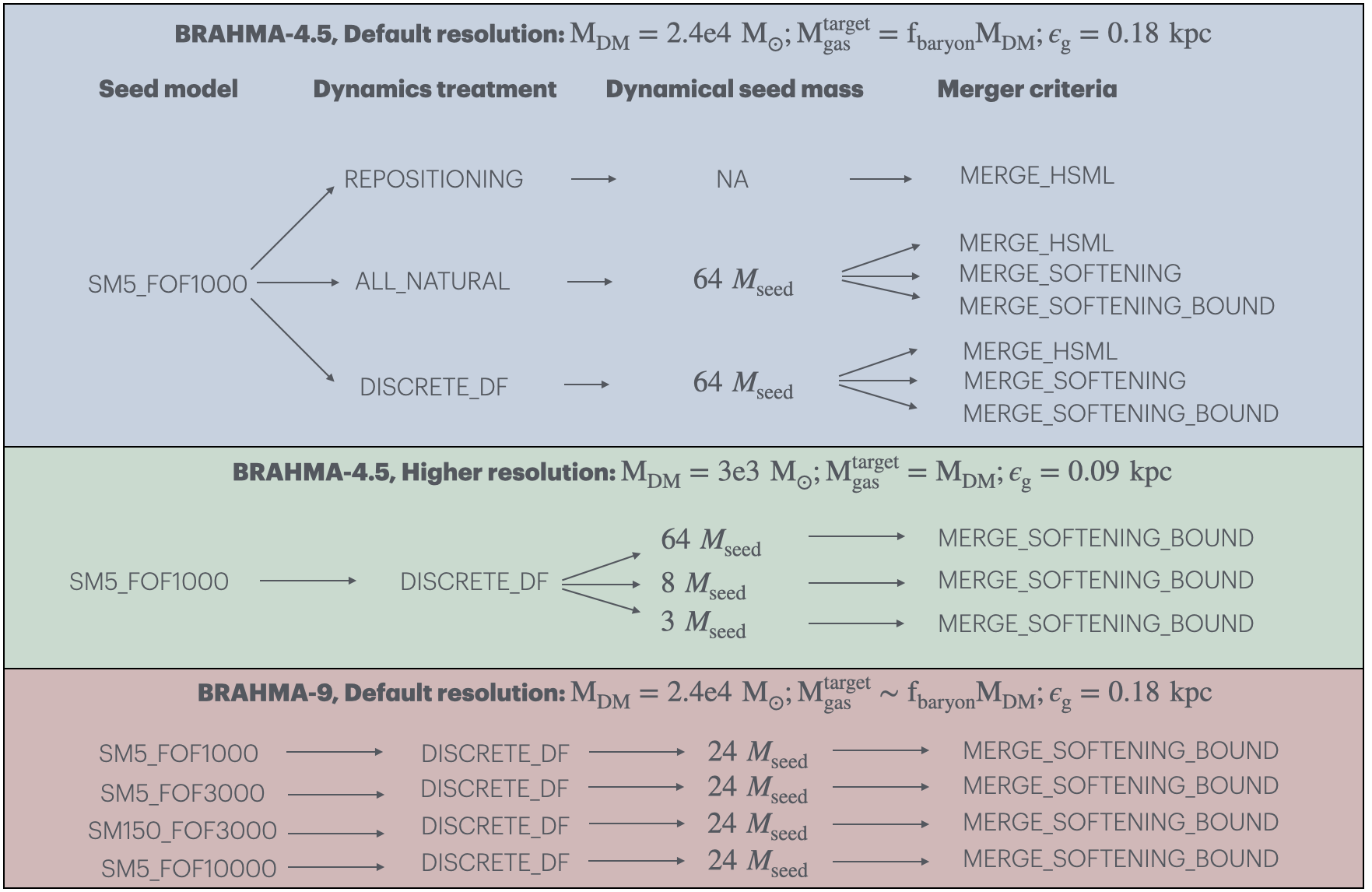}
\caption{Full summary of the simulation suite: Between the different seed models, dynamics treatments, dynamical seed masses and merging criteria, there are a total of 14 simulation boxes. The top row shows the default resolution $[4.5 \rm~Mpc]^3$ boxes which are used to quantify the impact of different dynamics treatments and merging criteria for a fixed seed model and dynamical seed mass. \texttt{REPOSITIONING} refers to boxes that use BH repositioning. \texttt{DISCRETE_DF} boxes remove BH repositioning and use the M23 subgrid DF model. The \texttt{ALL_NATURAL} boxes use neither BH repositioning nor a subgrid DF model, and simply follow the natural~(softened) Newtonian dynamics.  The middle row shows the higher resolution $[4.5~\rm Mpc]^3$~\texttt{DISCRETE_DF} boxes that are used to quantify the impact of different dynamical seed masses. Finally, the third row shows the $[9~\rm Mpc]^3$ default resolution \texttt{DISCRETE_DF} boxes that are used to show merger rate predictions for different seed models. In all the figures hereafter, plots that use the higher resolution boxes will have an explicit "Higher resolution" label on the legend or the captions. Plots that do not explicitly state the resolution, are made using default resolution boxes. }
\label{Summary_of_simulation_suite}
\end{figure*}
In this paper we present a new suite of simulations that extend the \texttt{BRAHMA} simulation suite~\citep{2024MNRAS.529.3768B,2024MNRAS.533.1907B} using the \texttt{AREPO} gravity + magneto-hydrodynamics~(MHD) code~\citep{2010MNRAS.401..791S,2011MNRAS.418.1392P,2016MNRAS.462.2603P,2020ApJS..248...32W}. 
The code solves for N-body gravity using the PM Tree~\citep{1986Natur.324..446B} method. 
The MHD is solved over a dynamic unstructured grid generated via a Voronoi tessellation of the domain.  The initial conditions were generated using \texttt{MUSIC}~\citep{2011MNRAS.415.2101H}. 
We adopt the \cite{2016A&A...594A..13P} cosmology i.e. $\Omega_{\Lambda}=0.6911, \Omega_m=0.3089, \Omega_b=0.0486, H_0=67.74~\mathrm{km}~\mathrm{sec}^{-1}\mathrm{Mpc}^{-1},\sigma_8=0.8159,n_s=0.9667$. 

To explore different BH dynamics treatments, we primarily used $[4.5~\mathrm{Mpc}]^3$ boxes that are referred to as \texttt{BRAHMA-4.5} following the convention from \citep{2024MNRAS.531.4311B}. The majority of these boxes have $512^3$ dark matter~(DM) particles and equal number of gas cells; the DM mass resolution is $2.4\times10^4~M_{\odot}$ for DM and the target gas mass resolution~($M^{\rm target}_{\rm gas}$) is $3.8\times10^3~M_{\odot}$ with gas cells subject to refinement and de-refinement to meet this target. These simulations have a gravitational softening length of $0.18~\mathrm{kpc}$ and we refer to them as ``default resolution boxes" for reasons that follow. 

For a select few of the \texttt{BRAHMA-4.5} boxes, we used a higher number of $1024^3$ DM particles to attain DM resolutions of $3\times10^3~M_{\odot}$ that are closer to the BH seed mass. These simulations have a gravitational softening length of $0.09~\mathrm{kpc}$. By default, \texttt{AREPO} generates gas cells 
which results in a target resolution of $4.7\times10^2~M_{\odot}$ for this setup. But to lower the computational expense, and to avoid the regime wherein the star particle masses would be comparable to individual massive stars, we reduced the number of gas cells and enforce the target  gas mass resolution to be equal to the DM resolution. This is done by globally de-refining the gas cells from their original resolution at the initial redshift. In principle, this ``global de-refinement" could negatively impact our results by inducing some noise in the initial power spectrum at the smallest scales. However, in Appendix \ref{appendix}, we determined that for a fixed seeding and dynamics model, the higher resolution simulations predict only $\sim2$ times fewer seeding rates and merger rates compared to the default resolutions. While this difference is not negligible, it is still small enough that we can use a handful of these boxes to test the impact of different model assumptions~(more details in Section \ref{Preventing numerical heating: Enhanced dynamical masses for the seed BHs}). We refer to these runs as ``higher resolution boxes".  

After performing an extensive exploration of the different dynamics treatments using the \texttt{BRAHMA-4.5} boxes, we choose a fiducial dynamics model~(which includes subgrid DF) and run larger volume~($[9~\mathrm{Mpc}]^3$ with $2\times1024^3$ DM particles and initial gas cells) \texttt{BRAHMA-9} boxes with the default resolution setup. We used these boxes to make final predictions for the BH-BH merger rates with different seed models. We also compare our predictions with the \texttt{BRAHMA-9} boxes from \cite{2024MNRAS.531.4311B} that used BH repositioning.

Most features of the galaxy formation model used in the \texttt{BRAHMA} simulations were inherited from the \texttt{IllustrisTNG} model~\citep{2017MNRAS.465.3291W,2018MNRAS.473.4077P}, which itself was based on the \texttt{Illustris} model~\citep{2013MNRAS.436.3031V, 2014MNRAS.438.1985T}. Primordial gas cooling rates from several species~(from $\mathrm{H},\mathrm{H}^{+},\mathrm{He},\mathrm{He}^{+},\mathrm{He}^{++}$) are calculated based on \cite{1996ApJS..105...19K}. Metal cooling rates are interpolated from pre-calculated tables as in \cite{2008MNRAS.385.1443S}. 
Gas heating can occur due to the presence of a uniform time-dependent UV background. Gas cells become star forming when their densities exceed $0.13~\mathrm{cm}^{-3}$. The star-forming gas cells represent an unresolved multiphase interstellar medium described by an effective equation of state~\citep{2003MNRAS.339..289S,2014MNRAS.444.1518V}. These gas cells also create star particles that represent single stellar populations~(SSPs) that are characterised by their age, metallicity and an initial mass function adopted from \cite{2003PASP..115..763C}. The subsequent stellar evolution for the SSPs are based on \cite{2013MNRAS.436.3031V} with modifications for \texttt{IllustrisTNG} as described in \cite{2018MNRAS.473.4077P}. This leads to chemical enrichment of the SSPs, which follows the evolution of seven species of metals~(C, N, O, Ne, Mg, Si, Fe) in addition to H and He. The enrichment of gas occurs when stellar and Type Ia/II supernova feedback, modeled as galactic scale winds~\citep{2018MNRAS.475..648P}, ejects the metals from the SSPs into the surrounding gas.

Gas accretion on to BHs is modeled by the Eddington-limited Bondi-Hoyle formalism described as 
\begin{eqnarray}
\dot{M}_{\mathrm{bh}}=\mathrm{min}(\dot{M}_{\mathrm{Bondi}}, \dot{M}_{\mathrm{Edd}})\\
\dot{M}_{\mathrm{Bondi}}=\frac{4 \pi G^2 M_{\mathrm{bh}}^2 \rho}{c_s^3}\\
\dot{M}_{\mathrm{Edd}}=\frac{4\pi G M_{\mathrm{bh}} m_p}{\epsilon_r \sigma_T~c}
\label{bondi_eqn}
\end{eqnarray} 
where $G$ is the gravitational constant, $\rho$ is the local gas density, $M_{\mathrm{bh}}$ is the BH mass, $c_s$ is the local sound speed, $m_p$ is the proton mass, and $\sigma_T$ is the Thompson scattering cross section. Accreting BHs radiate with bolometric luminosities given by $L_{\mathrm{bol}}=\epsilon_r \dot{M}_{\mathrm{bh}} c^2$
where the radiative efficiency $\epsilon_r$ is assumed to be 0.2. AGN feedback is implemented via two distinct modes depending on the Eddington ratio $\eta$. `Thermal feedback' is implemented for high Eddington ratios~($\eta > \eta_{\mathrm{crit}}\equiv\mathrm{min}[0.002(M_{\mathrm{BH}}/10^8 \mathrm{M_{\odot}})^2,0.1]$) wherein a fraction of the radiated luminosity is deposited to the gas. `Kinetic feedback' is injected for low Eddington ratios~($\eta < \eta_{\mathrm{crit}}$) at irregular time intervals along a randomly chosen direction. As we shall see, BHs do not undergo substantial accretion-driven growth in our simulations, as the accretion of gas on to low mass seeds in low mass halos is suppressed by supernova feedback as well as the $M_{bh}^2$ scaling of the Bondi accretion model. Therefore, AGN feedback is not consequentual to our results. Nevertheless, readers interested in further details about the AGN feedback modeling and its impact on \texttt{IllustrisTNG} galaxies can refer to \cite{2017MNRAS.465.3291W}, \cite{2019MNRAS.490.3234N} and \cite{2020MNRAS.493.1888T}. 

\subsection{Black hole seeding}
\label{Black hole seeding}
We use the gas based BH seed models built in previous work~\citep{2021MNRAS.507.2012B} summarized as follows: Our seed mass is $\seedmass=2.2\times10^3~M_{\odot}$, which is close to the gas mass resolution of the simulation. We place these seeds in halos that exceed critical thresholds of star-forming \& metal-poor gas mass and halo mass, denoted by $\msfmp$~(``sf" stands for ``star-forming" and ``mp" stands for ``metal-poor") and $\mh$ respectively in the units of $\seedmass$. Recall that star-formation occurs for gas cells with densities $>0.13~\rm cm^{-3}$. The densest, metal poor gas cell inside the halos is then converted to a BH seed. In \cite{2021MNRAS.507.2012B}, we systematically explored a range of seed models with different combinations of values adopted for $\msfmp$~($=5,50~\&~150$) and $\mh$~($=1000,3000~\&~10000$) using zoom simulations. 
These seed models converge well for formation rates with gas mass resolutions $\lesssim10^4~M_{\odot}$. 

Following \cite{2024MNRAS.529.3768B}, we label our seed models as \texttt{SM*_FOF*} where the two `*'s correspond to values for $\msfmp$ and $\mh$ respectively. Throughout most of the paper, we adopt a single seed model, namely \texttt{SM5_FOF1000}~($\msfmp,\mh=5,1000$) and examine various dynamical scenarios using \texttt{BRAHMA-4.5} boxes. In the latter sections, we pick a fiducial dynamics model and use \texttt{BRAHMA-9} boxes to present BH merger rate predictions for different seed model variations \texttt{SM5_FOF3000} ($\msfmp,\mh=5,3000$), \texttt{SM150_FOF3000} ($\msfmp,\mh=150,3000$), and \texttt{SM5_FOF10000} ($\msfmp,\mh=5,10000$).

\subsection{Black hole dynamics}
\label{Black hole dynamics}
\subsubsection{Accounting for missing BH dynamical friction from background particles}
\label{DF_model}
To accurately model small-scale BH dynamics, we correct for the missing DF force that would arise with sufficient spatial and mass resolution. In this work, we apply the DF estimator built by M23 for discrete N-body systems 
given by 
\begin{equation}
\textbf{a}_{\rm df} = \sum_i \frac{\alpha_i b_i}{(1+\alpha_i^2)r_i} \left(S_i(r_i)\frac{G \Delta m_i}{r_i^2}\right) \hat{\textbf{V}}_i 
\end{equation}
with $\alpha_i \approx b_i V_i^2/G M_{bh}^{i}$ and $b_i\equiv r_i \left| \hat{\textbf{r}}_i - (\hat{\textbf{r}}_i.\hat{\textbf{V}}_i) \hat{\textbf{V}}_i \right|$. The factor $S_i$ corresponds to the gravitational softening scheme used in their simulations. In our simulations, the gravitational softening is implemented by replacing $r_i$ by $r_i+r_{\mathrm{soft}}$ where $r_{\mathrm{soft}}$ is softening length for a given particle type. Therefore, as applied to our simulation setup, the formula changes to 
\begin{equation}
\textbf{a}_{\rm df} = \sum_i \frac{\alpha_i b_i}{(1+\alpha_i^2)(r_i+r_{\rm soft})} \left(\frac{G \Delta m_i}{(r_i+r_{\rm soft})^2}\right) \hat{\textbf{V}}_i.
\label{DF_eqn}
\end{equation} 

Aside from the BH mass $M_{bh}$, the formula only depends on $\Delta m_i$, $\textbf{r}_i$; and $\textbf{V}_i$ corresponding to the mass; relative displacement and relative velocity of each resolution element with respect to the BH. The summation is performed over all the mass resolution elements that the code encounters on the PM tree during the usual gravity calculation. In order to manifestly conserve momentum, the prescription also applies a `back reaction' acceleration, $-\frac{\Delta m_i}{M_{bh}}\Delta a_i$, on the mass resolution elements. 

As highlighted by M23, their DF model has the distinct advantage of not having any free parameters or assumptions of isotropic and homogeneous distributions of particles, as opposed to several models that are based on the \cite{1943ApJ....97..255C} formalism~\citep{2015MNRAS.451.1868T,2019MNRAS.486..101P,2022MNRAS.510..531C,2024MNRAS.534..957G}. In addition, this prescription manifestly conserves the total momentum, which is difficult to do with most of the other prescriptions. Finally, the M23 expression exactly reduces to the Chandrashekhar DF formula for the relevant assumptions wherein the background particles assume a time invariant, spatially homogeneous distribution and a Maxwellian velocity distribution. To that end, the core strength of the M23 treatment is that it generally holds for a wide range of arbitrary phase space distributions of background particles that are expected in cosmological simulations.  

The simulation set that use our implementation of the M23 subgrid DF model are labeled as \texttt{DISCRETE_DF}. Additionally, we also conduct simulations using the traditional repositioning method (\texttt{REPOSITIONING}) and an alternative approach where BHs follow purely Newtonian dynamics without repositioning or subgrid DF, referred to as \texttt{ALL_NATURAL}.

\subsubsection{Preventing numerical heating: Enhanced dynamical masses for the seed BHs}
\label{Preventing numerical heating: Enhanced dynamical masses for the seed BHs}

Due to numerical heating, nearly all the subgrid DF estimators~(including the M23 prescription used here) break down in the regime when the BH mass is close to or smaller than the background particle mass. \cite{2024MNRAS.534..957G} is the only model that attempts to correct for the impact of this numerical heating for BH masses between $\sim1-5$ times the background particle mass, but they are yet to explore the regime where the BH mass is less than the background particle mass in the simulation box. In many cosmological simulations, the seed mass is below the DM particle mass. In such cases, large scale simulations that use subgrid DF~\citep[e.g.][]{2022MNRAS.513..670N} typically assume an enhanced ``dynamical" seed mass~($\seedmassdynamical$) for the gravity calculation, while the ``actual" seed mass~($\seedmass$) is used for BH accretion and AGN feedback. 

For our simulations, we explore $\seedmassdynamical\sim3-64~\times \seedmass$, which ensures that it is always at least $\sim3$ times the DM particle mass. Notably, M23 reports that their model works best when the BH masses are $\sim10$ times the background particle masses. We start with a value close to that choice with $\seedmassdynamical =  64~\seedmass \sim 6~M_{\rm DM}$ in the default resolution boxes with $M_{\rm DM} = 2.4\times10^4~M_{\odot}$, to test the impact of different dynamical treatments. Upon confirming that $\seedmassdynamical =  64~\seedmass$ leads to effective sinking and merging of BHs in the presence of subgrid DF, we also test models with lower values of the $\seedmassdynamical$ even though they are significantly below the M23 recommendation. This is where we need the higher resolution boxes with $1024^3$ DM particles to close the gap~(as much as possible) between $\seedmassdynamical$ and $\seedmass$ while still mitigating the impact of numerical heating. As we shall see later, our assumed $\seedmassdynamical$ values are large enough to ensure that a significant portion of BHs effectively sink to the halo centers and participate in mergers when subgrid DF is applied. However, reducing the dynamical masses does have a significant impact on the BH merger rates, which we shall quantify in Section \ref{Impact of dynamical seed mass}.      
\subsubsection{Black hole mergers}
\label{mergers}
 
Adding a subgrid DF model allows us to trace the evolution of merging BH pairs to smaller separations compared to repositioning, close to the gravitational softening length. As a result, simulations such as \texttt{ASTRID}, \texttt{ROMULUS} and \texttt{OBELISK}~(that use subgrid DF) have used minimum merger separations of a few~($\sim2-4$) times the gravitational softening length. Subgrid DF also allows us to explicitly trace the energy dissipation of the BH pairs to the point where they finally become gravitationally bound. This has allowed several simulations~(e.g. \texttt{ASTRID} and \texttt{ROMULUS}) to also apply the requirement for the merging BH pairs to be gravitationally bound.

In this work, we test the impact of different sets of merging criteria on the merger rates. These are as follows:
\begin{itemize}
\item \texttt{MERGE_HSML}: We merge BH pairs solely based on their separation distances, particularly when BH pair separations are below their neighbor search radii~($R_{\rm hsml}\sim1-15 \rm~kpc$). This criterion was used in the original \texttt{BRAHMA} simulations, inherited from the \texttt{Illustris-TNG} simulations.
\item \texttt{MERGE_SOFTENING}: 
We merge BH pairs when 
their separation distances are 
less than 2 times the gravitational softening length. For the lower resolution and higher resolution simulations, this corresponds to merging separations of $0.18\rm~kpc$ and $0.36\rm~kpc$ respectively.

\item \texttt{MERGE_SOFTENING_BOUND}: In addition to requiring the pair separations to be less than 2 times the gravitational softening, we also require the BH pairs to be gravitationally bound. Specifically, we require $\bf{|\Delta v|^2/2 +  \Delta a.\Delta r} < 0$ where $\bf{\Delta v}$, $\bf{\Delta a}$ and $\bf{\Delta r}$ are the relative velocities, relative accelerations and relative displacement between the BH pairs. 

\end{itemize}

We consider \texttt{MERGE_SOFTENING_BOUND} to be 
our fiducial merger criteria, but 
we also include simulations with \texttt{MERGE_HSML} and \texttt{MERGE_SOFTENING} to quantify the impact of different components of the merger criteria~(separation distances and gravitational boundedness). The \texttt{MERGE_SOFTENING_BOUND} criteria was first adopted by \cite{2011ApJ...742...13B}, and has now become the standard criteria used in many recent cosmological simulations~\citep{2017MNRAS.470.1121T,2021MNRAS.505.5129B,2022MNRAS.513..670N,2022MNRAS.510..531C}. Since in our simulations, the \texttt{MERGE_SOFTENING_BOUND} criteria merges bound BH pairs at $0.18$ and $0.36$ kpc depending on the resolution,  we phrase this as ``tracking the evolution of BH pairs down to $\sim0.2-0.4\rm~kpc$ scales" throughout the paper.

\subsection{Full summary of our simulation suite and nomenclature}
\label{Full summary of our simulation suite and nomenclature}



Our simulation consists of multiple boxes - \texttt{BRAHMA-4.5} and \texttt{BRAHMA-9} designed to systematically explore the impact of different choices for BH seeding, dynamics, and mergers. We have four seed model variations~(\texttt{SM5_FOF1000}, \texttt{SM5_FOF3000}, \texttt{SM150_FOF3000} and \texttt{SM5_FOF10000}), three dynamics model variations~(\texttt{ALL_NATURAL}, \texttt{DISCRETE_DF} and \texttt{REPOSITIONING}), four values for dynamical seed masses~($\seedmassdynamical = 3,8,24~\&~64~\seedmass$), and three different merger criteria~(\texttt{MERGE_HSML}, \texttt{MERGE_SOFTENING}, and \texttt{MERGE_SOFTENING_BOUND}). Furthermore, we have two configurations of mass resolution for dark matter particles (DM) and gas cells:  ``default resolution"~($M^{\rm target}_{\rm gas} =  f_{\rm baryon}~M_{\rm DM}; M_{\rm DM} = 2.4\times10^4~M_{\odot}$) and ``higher resolution"~($M^{\rm target}_{\rm gas} = M_{\rm DM} = 3\times10^{3}~M_{\odot}$). 


Overall, each simulation box is characterized by six key features: the seed model, dynamics model, initial dynamical mass for seeds, merger criteria, volume~(\texttt{BRAHMA-4.5} vs \texttt{BRAHMA-9}) and resolution level (default vs. higher resolution). With different combinations of these features, we generate 14 simulation boxes, summarized in Figure \ref{Summary_of_simulation_suite}, which are grouped into three categories.

The first category, shown in the top row of Figure \ref{Summary_of_simulation_suite}, consists of seven default resolution \texttt{BRAHMA-4.5} simulations  that adopt a fixed seed model (\texttt{SM5_FOF1000}) and a fixed dynamical seed mass ($\seedmassdynamical = 64~\seedmass$). These simulations explore different combinations of dynamical treatments (\texttt{REPOSITIONING}, \texttt{ALL_NATURAL}, and \texttt{DF_DISCRETE}) and merger criteria (\texttt{MERGE_HSML}, \texttt{MERGE_SOFTENING}, and \texttt{MERGE_SOFTENING_BOUND}). These runs assess the impact of various BH dynamics treatments on merger rates.

The second category, presented in the middle row of Figure \ref{Summary_of_simulation_suite}, includes three higher resolution \texttt{BRAHMA-4.5} simulations that maintain a fixed seed model (\texttt{SM5_FOF1000}), dynamics model (\texttt{DF_DISCRETE}), and merger criteria (\texttt{MERGE_SOFTENING_BOUND}) while varying the dynamical seed mass. These simulations are designed to quantify how the dynamical seed mass affects BH merger rates, down to $\seedmassdynamical = 3 ~\seedmass$.

Finally, the third category, displayed in the bottom row of Figure \ref{Summary_of_simulation_suite}, consists of four default resolution \texttt{BRAHMA-9} simulations. These adopt the fiducial dynamics model (\texttt{DF_DISCRETE}), merger criteria (\texttt{MERGE_SOFTENING_BOUND}), and a fixed dynamical seed mass ($\seedmassdynamical = 24~\seedmass$) while varying the seed model. These runs provide final BH merger rate predictions across different seed models.

\section{Results}
\label{results}
\subsection{Seed formation}
\label{Seed formation section}
\begin{figure*}
\includegraphics[width= 5.9cm]{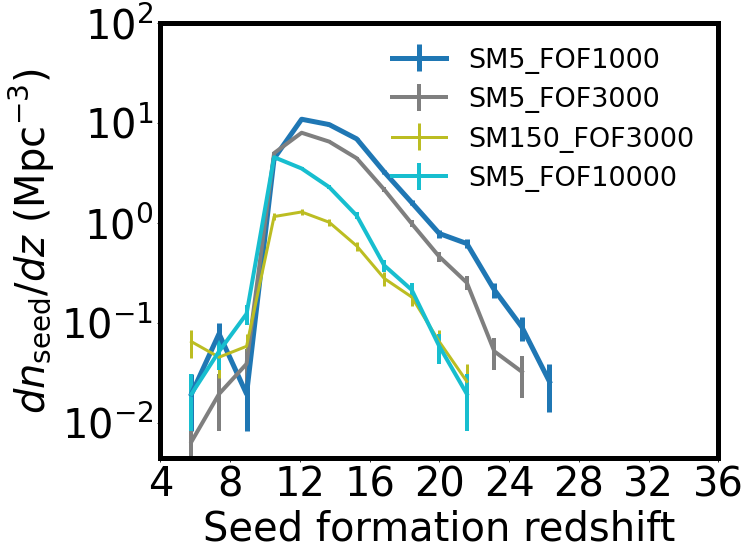}
\includegraphics[width= 5.6cm]{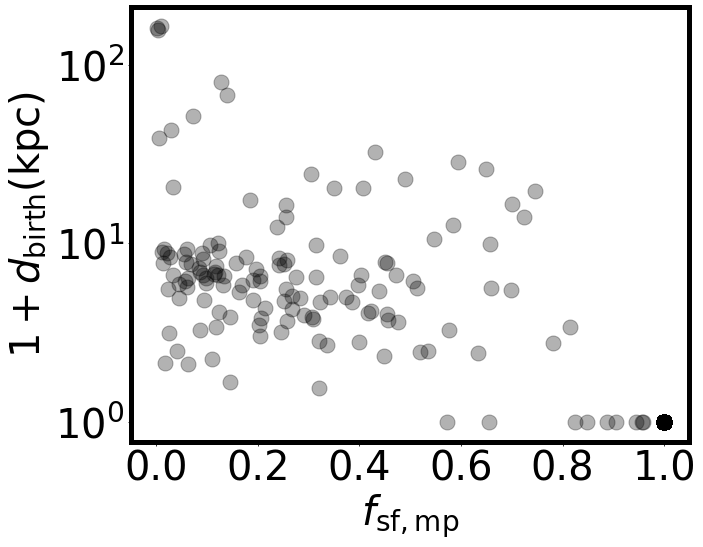}
\includegraphics[width= 6.1cm]{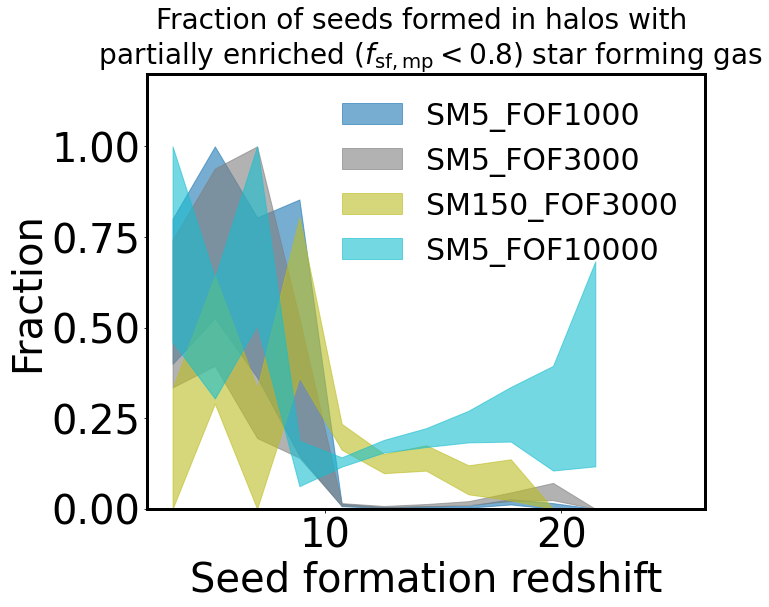}

\caption{Leftmost panel shows the comoving number density of new seeds formed at different redshifts, for four different boxes assuming different seed models. The middle panel shows the distance $d_{\rm birth}$ between the seed birth site and the halo center (the densest gas cell), plotted against the metal-poor fraction of star-forming gas in the halo, $f_{\rm sf,mp} \equiv M_{\rm sf,mp}/M_{\rm sf}$, for seed-forming halos ($\msfmp, \mh = 5, 10000$) at redshifts $z \sim 5\text{--}20$.
For those seed-forming halos wherein the star-forming gas is \textit{partially metal enriched}~(here defined as $f_{\rm sf,mp} < 0.8$ and referred to as partially enriched halos), seeds typically form at distances of $\sim2-100~\rm kpc$ from the halo center. The right most panel shows the overall fraction of seeds that form in partially enriched halos. The majority of seeds do not form in partially enriched halos, especially at $z\gtrsim10$. Therefore, most seeds form very close to the halo center~(at the densest gas cell of the halo).}
\label{seed formation}
\end{figure*}
The left panel of Figure \ref{seed formation} shows the formation rates of new seeds as a function of redshift for the four \texttt{BRAHMA-9} boxes. 
Seed formation starts around $\sim20-25$, 
peaks at $z\sim10$, and substantially drops thereafter due to metal enrichment and a reduction in gas content in low-mass halos. 
The latter can be attributed to stellar feedback as well as cosmological expansion, causing halos of fixed mass to become less bound at later redshifts. 


The birth location of 
seeds 
within halos is expected to influence their dynamical fates. 
Specifically, our seeds are more likely to be at the halo centers~(and merge with other seeds) if they are born there. 
Whether the seed is formed at the halo center or not is essentially determined by whether its densest gas cell~(defined as the halo center) is also metal-poor. If not, the seed would be formed at the \textit{densest metal-poor} gas cell at some distance away from the \textit{densest}~(but not metal-poor) gas cell of the halo. 

The middle panel of Figure \ref{seed formation} shows the distance~($d_{\rm birth}$) between the seed birth site and the halo center as a function of the star-forming \& metal-poor gas mass fraction $f_{\rm sf,mp}\equiv M_{\rm sf,mp}/M_{\rm sf}$, for seed formation sites spanning $z=5-20$. Here, $M_{\rm sf,mp}$ and $M_{\rm sf}$ are the total star-forming \& metal-poor gas mass and the total star-forming gas mass respectively, in a given halo~\footnote{Note the distinction between $M_{\rm sf,mp}$ and $\tilde{M}_{\rm sfmp}$. The former, $M_{\rm sf,mp}$, denotes an actual halo property, while the latter, $\tilde{M}_{\rm sfmp}$, is a seeding threshold applied based on that property. More precisely, the \textit{dense and metal-poor gas mass criterion} seeds BHs in halos where $M_{\rm sf,mp} > \tilde{M}_{\rm sfmp}~\seedmass$.
}. We can naturally see that $d_{\rm birth}$ decreases when $f_{\rm sf,mp}$ increases. In other words, in more enriched halos, seeds are more likely to form in pockets of dense, metal-poor gas away from the halo center. When most of the star-forming gas inside the halo is metal-poor~($f_{\rm sf,mp}> 0.8$), then the birth site is almost always at the halo center. On the other hand, for seed forming halos wherein less than $\sim80\%$ of the star-forming gas is metal-poor~(defined here as $f_{\rm sf,mp}< 0.8$), the birth sites are at distances ranging from $\sim2-20~\mathrm{kpc}$ from the halo center. Hereafter, we refer to halos with $f_{\rm sf,mp}< 0.8$ \textit{partially enriched halos}.

The rightmost panel of Figure \ref{seed formation}  shows the relative fraction of seeds that form inside partially enriched halos. In all the seed models, we can generally see that the fraction of seeds forming within partially enriched star-forming gas sharply increases at $z\lesssim10$, which unsurprisingly coincides with the suppression of seed formation by metal enrichment at these redshifts. Let us now look at $z\gtrsim10$ wherein most seeds form. For the two most lenient seed models~(\texttt{SM5_FOF1000} and \texttt{SM5_FOF3000}), only a very small fraction~(less than a few percent) of seeds form in halos with partially enriched star-forming gas. For the stricter seed models with higher mass thresholds~(\texttt{SM150_FOF3000} and \texttt{SM5_FOF10000}), a  higher fraction~($\sim15-40~\%$) of seeds form in halos with partially enriched star-forming gas. This is not surprising since in these models, some of the star-forming gas is likely to be enriched by the time the mass threshold is reached. Nevertheless, in all the seed models, the majority of seeds do not form in halos with partially enriched star-forming gas. More precisely, most seeds form in halos wherein $\gtrsim80~\%$ of the star-forming gas is metal poor at the time of seed-formation~(although they are rapidly metal enriched soon after seed formation,~\citealt{2024MNRAS.529.3768B}). Therefore, the majority of the seeds form very close to the halo center~(i.e. location of the densest gas cell in the halo).

Overall, since most of our seeds form close to the halo center, they have a better chance of participating in BH-BH mergers after their host halos merge. While they could still migrate away from the halo center due to dynamical interactions and GW recoils~(which we do not explicitly include here), being born at the halo center still provides the best prospect for BHs to encounter and merge with other BHs after the halos merge.  


\subsection{Evolution of black hole abundances}
\begin{figure}
\noindent

\includegraphics[width= 4.7cm]{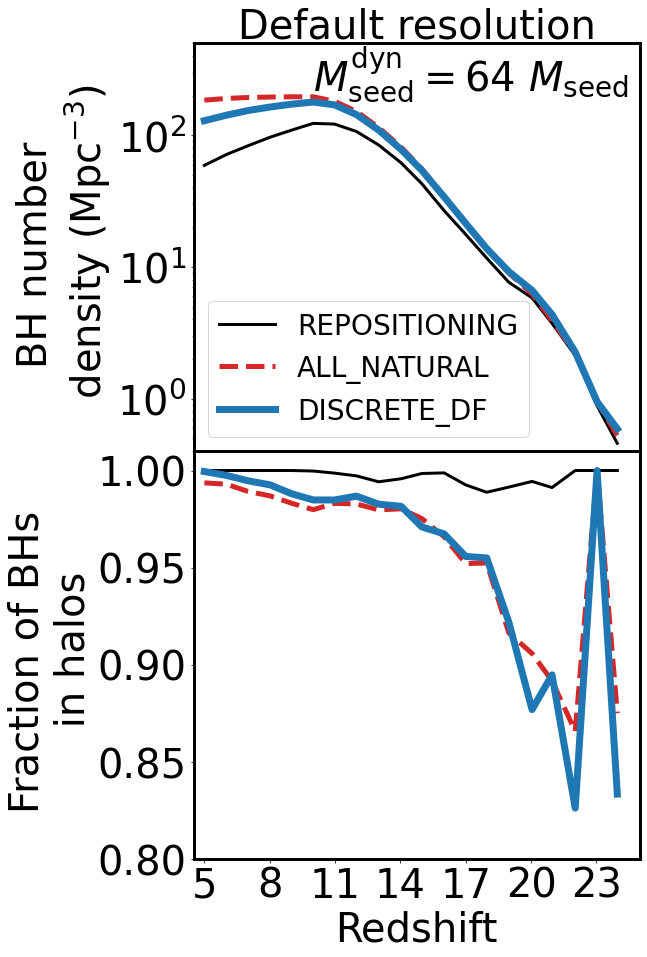} \hspace{-1.7mm}\includegraphics[width= 3.37cm]{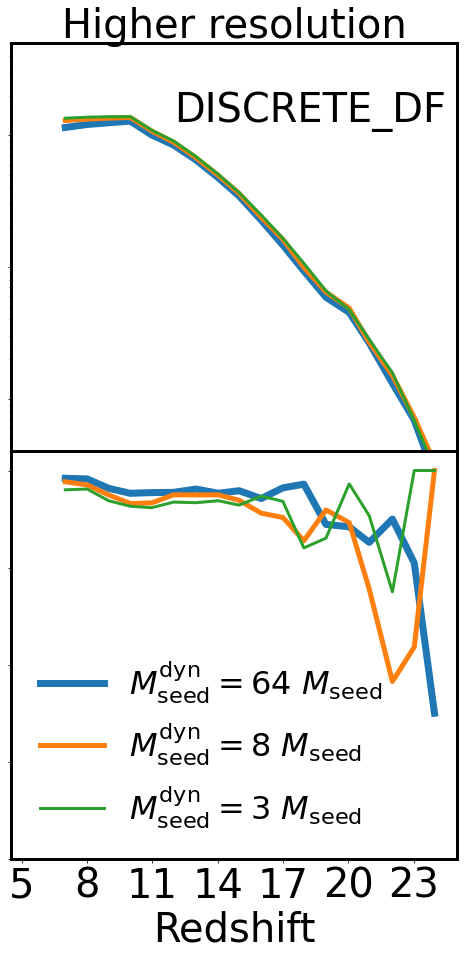}
\caption{
The 
number density of BHs (top) and the fraction of BHs residing inside halos (bottom) as a function of redshift using the \texttt{SM5_FOF1000} seed model.
Each line indicates a different BH dynamics model (left) or BH seed mass (right) as indicated in the legends.
Removing \texttt{REPOSITIONING} leads to $\sim15\%$ of BHs wandering outside halos at $z\gtrsim20$. However, by $z\sim7$, almost all BHs are found to be inside halos. Adding the subgrid DF does not have a significant impact on the population of BHs wandering outside halos, regardless of the dynamical seed masses.}
\label{BH_wandering}
\end{figure}
We now look at the overall evolution of the BH number densities in our simulations under different treatments for BH dynamics. Note that hereafter, for the majority of the paper, we shall focus on one of the seed models, namely the \texttt{SM5_FOF1000} model, and compare different dynamics treatments described in Section \ref{Black hole dynamics} using the \texttt{BRAHMA-4.5} boxes. In Section \ref{Merger rates for different seed models}, we will revisit the other seed models and look at how the BH merger rate predictions are impacted by BH dynamics.  

The BH number density evolution is shown in the top panels of Figure \ref{BH_wandering}. From $z\sim20$ to $z\sim10$, number density naturally increases with time due to the formation of new seeds. In this regime, the dynamics treatment is not very consequential to the overall number density. At $z\lesssim10$ when seed formation is suppressed, the number density evolution is driven by the prevalence of BH-BH mergers. This is naturally the regime wherein the different dynamics treatments start to more significantly impact the overall number densities. When repositioning is applied~(black line in the left panel), the BHs are promptly merged as their host halos merge. As a result, the number density starts to decrease significantly faster than in the boxes that do not use repositioning. 
For the 
\texttt{ALL_NATURAL} box wherein neither repositioning nor subgrid DF is applied~(red dashed line in the left panel), the number density is almost constant at $z\lesssim10$. This is because the BH-BH mergers become very inefficient. 
Notably, here we apply the strictest criteria for mergers~(\texttt{MERGE_SOFTENING_BOUND}). 
For the \texttt{DISCRETE_DF} box where the subgrid DF model is applied~(solid blue line), the number density decreases with time at $z\lesssim10$, but not as rapidly as the \texttt{REPOSITIONING} box. 
Overall, these results hint that in the absence of repositioning, the subgrid DF model does play a crucial role in ensuring that our seeds participate in BH-BH mergers. 
However, the merger rate is lower compared to models that use repositioning. 

The bottom panels of Figure~\ref{BH_wandering} show the fraction of BHs associated with a host halo as a function of redshift. When repositioning is used, nearly $\sim100~\%$ of the BHs remain within halos. In contrast, in the \texttt{ALL_NATURAL} and \texttt{DISCRETE_DF} models, a small fraction of BHs are found to be not associated with any halo, particularly at higher redshifts~($\sim10$--$15~\%$ at $z\sim20$). Notably, this fraction is largely unaffected by the inclusion of subgrid DF~(bottom-left panel), and it also shows minimal dependence on the BH-to-DM particle mass ratio~(bottom-right panel). However, we find a slight reduction in the fraction of these ``orphaned'' BHs with increasing resolution~(compare bottom-left and bottom-right panels). This suggests that the presence of such BHs is primarily a resolution-dependent numerical artifact. More specifically, the limited resolution in low-mass halos, combined with numerical noise in the FOF halo finder—can lead to failures in associating BHs with their host halos. This issue is likely more prevalent when seeds form near the outskirts of halos. It may also explain why the \texttt{REPOSITIONING} runs do not produce orphaned BHs, since the seeds are re-located to the nearest potential minima, regardless of their initial formation site. Nevertheless, it is encouraging that the fraction of these orphaned BHs remains small and becomes negligible by $z\sim5$ in both the \texttt{ALL_NATURAL} and \texttt{DISCRETE_DF} boxes. Overall, this is consistent with expectations: since our simulations do not include GW-induced recoil kicks, we do not anticipate BHs becoming gravitationally unbound from their host halos, especially given that most seeds do form at halo centers. 


\begin{figure*}
\centering
\includegraphics[width= 14cm]{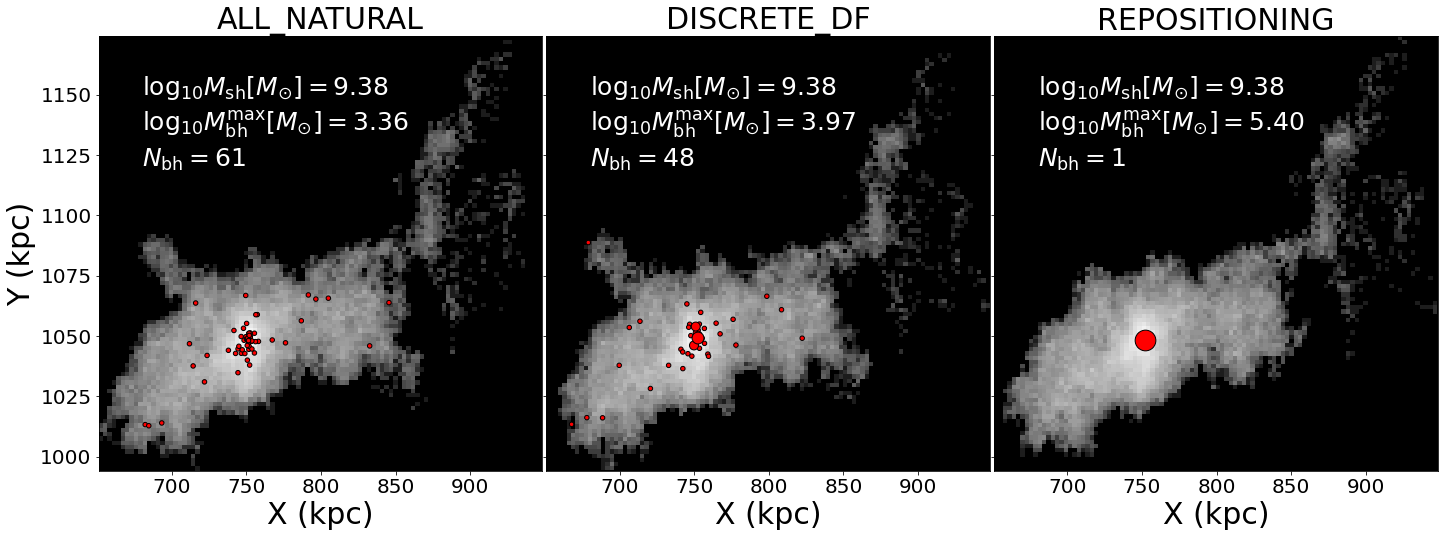}
\includegraphics[width= 14cm]{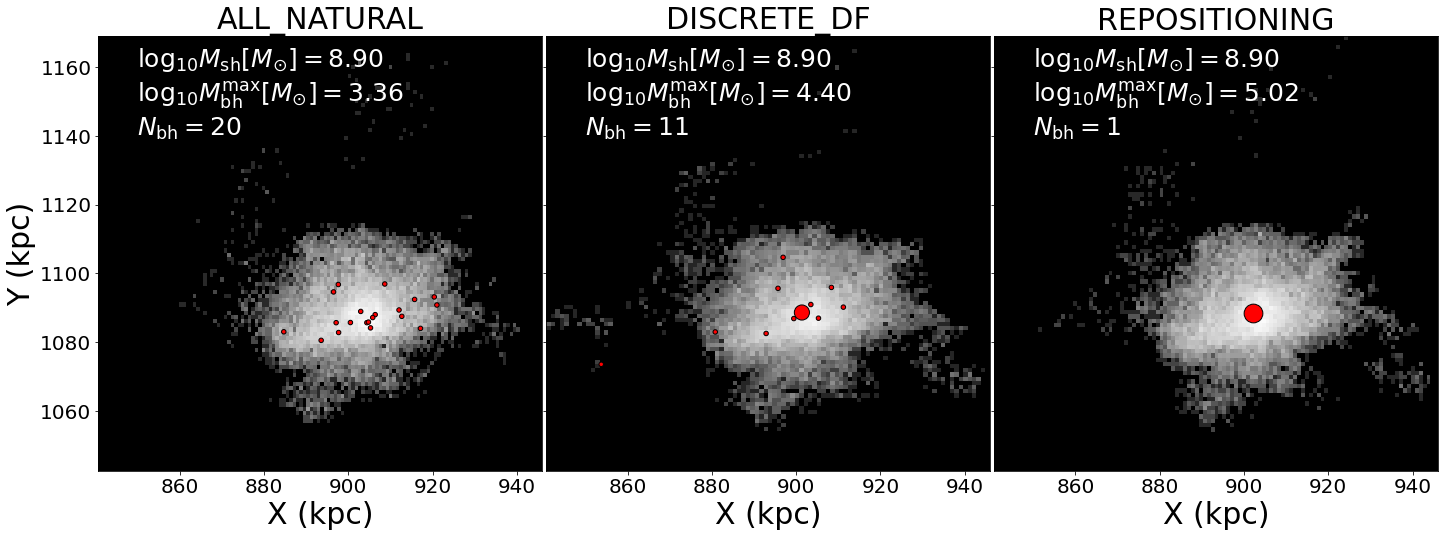}
\includegraphics[width= 14cm]{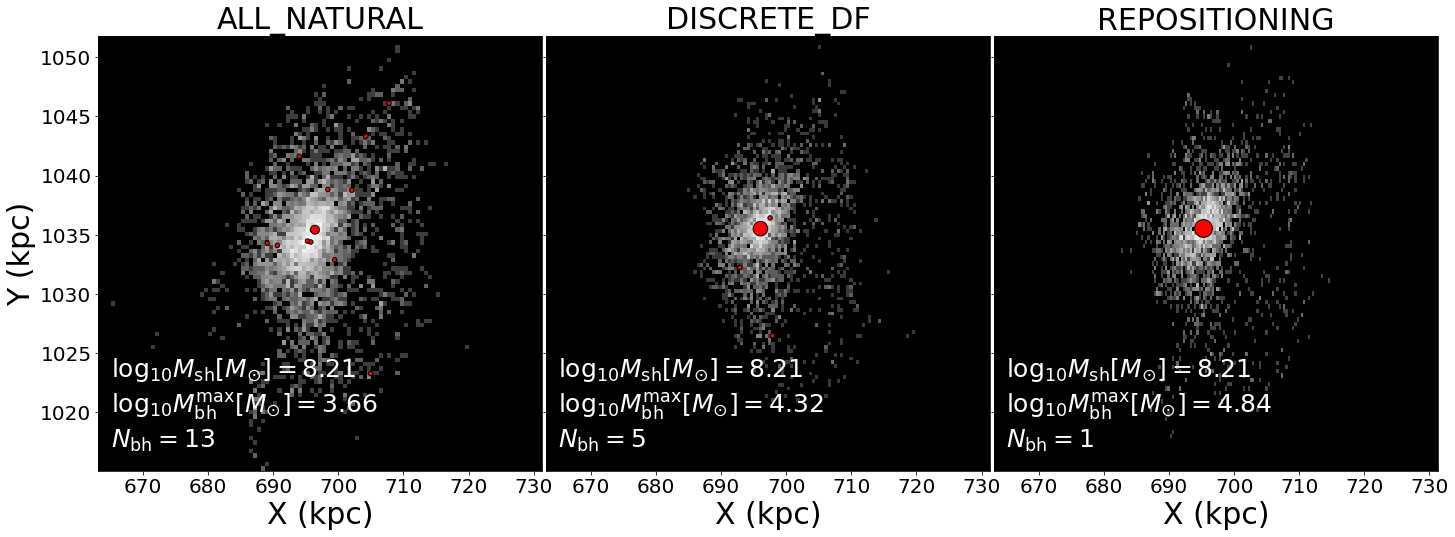}
\caption{The grey histograms show the DM distributions of a select few subhalos in our simulations~(at $z=9$). Their BHs are shown as red circles whose sizes are proportional to their masses~(in log units). Each column corresponds to a different treatment for BH dynamics, namely \texttt{ALL_NATURAL}, \texttt{DISCRETE_DF} and \texttt{REPOSITIONING}, from left to right. Top to bottom rows show three different subhalos of successively decreasing masses. When repositioning is applied, all the BHs merge to form a single central BH. Without repositioning, there are large numbers of secondary BHs in the subhalo that are yet to merge with the central BH. In the absence of repositioning, the subgrid DF correction is crucial to ensure that some mergers do occur.}
\label{BH_wandering_images}
\end{figure*}

\begin{figure}
\centering
\includegraphics[width= 7cm]{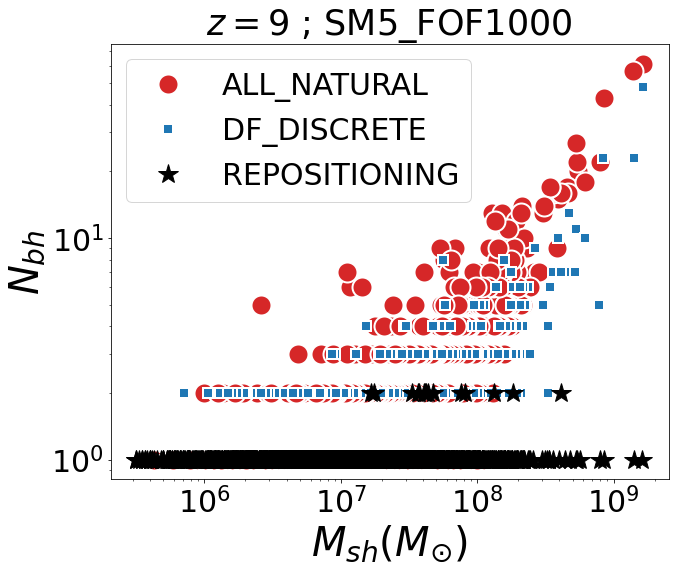}

\includegraphics[width= 7cm]{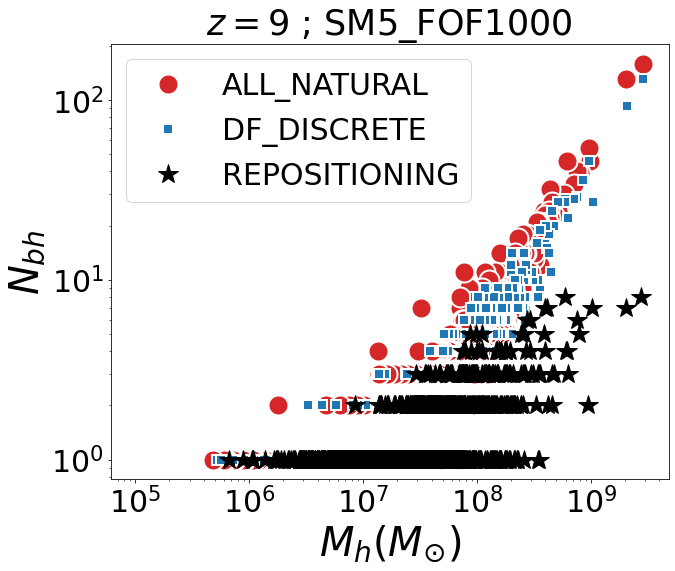}

\caption{BH multiplicity as a function of subhalo mass~(top panel) and halo mass~(bottom panel) at $z=9$ for three BH dynamics models.
The multiplicities increase when the merging efficiency decreases, with \texttt{REPOSITIONING} producing the lowest multiplicities and \texttt{ALL_NATURAL} producing the highest multiplicities.}
\label{BH_multiplicity}
\end{figure}

\subsection{Dynamics of BHs inside their host subhalos}

\label{Dynamics of BHs inside their host subhalos}
To examine the dynamics of BH seeds within their host subhalos, we visualize their distribution in an illustrative sample in Figure \ref{BH_wandering_images}.  We focus on three subhalos with masses ranging from $\sim10^8~M_{\odot}$ to a few times $\sim10^9~M_{\odot}$, which have grown substantially since their progenitor halos ($2.2\times10^6~M_{\odot}$) first formed the seeds. This evolution allows sufficient time for the dynamical effects to manifest. 
In the \texttt{REPOSITIONING} scheme~(rightmost panel), each subhalo contains only one BH, which has grown significantly ($\gtrsim10^5~M_{\odot}$). By contrast, the \texttt{ALL_NATURAL} scheme yields numerous BHs that have barely grown past their seed mass. In the most massive subhalo (top row), many BHs cluster near the subhalo center without merging -- likely owing to the strict \texttt{MERGE_SOFTENING_BOUND} criterion. The \texttt{DISCRETE_DF} scheme enables moderate BH growth ($\gtrsim10^4~M_{\odot}$), though significantly less than in \texttt{REPOSITIONING}, and results in fewer BHs compared to the \texttt{ALL_NATURAL} model.

\subsubsection{BH multiplicities} 

Figure \ref{BH_multiplicity} shows the BH multiplicty~(i.e. number of BHs) for the full population of subhalos~(top panel) and halos~(bottom panel) in our simulations. The \texttt{REPOSITIONING} runs yield the lowest multiplicities, with most halos containing only one BH. In this scheme, halo multiplicities greater than one arise only due to the presence of subhalos. The most massive $\sim10^9~M_{\odot}$ halos, with the highest number of subhalos, contain $\sim9$ BHs. To that end, almost all subhalos host a single BH, with only a small fraction containing an additional BH. 
Without repositioning, BH multiplicities increase significantly as mergers become less efficient. 
In the \texttt{ALL_NATURAL} simulations, the most massive halos and subhalos contain up to $\sim100$ and $\sim50$ BHs, respectively. The \texttt{DISCRETE_DF} scheme also reduces multiplicities by factors of $\sim2-4$ compared to \texttt{ALL_NATURAL}, as BHs in these runs merge more frequently.

\subsubsection{Spatial distributions of BHs inside subhalos}

\begin{figure*}
\includegraphics[width= 18cm]{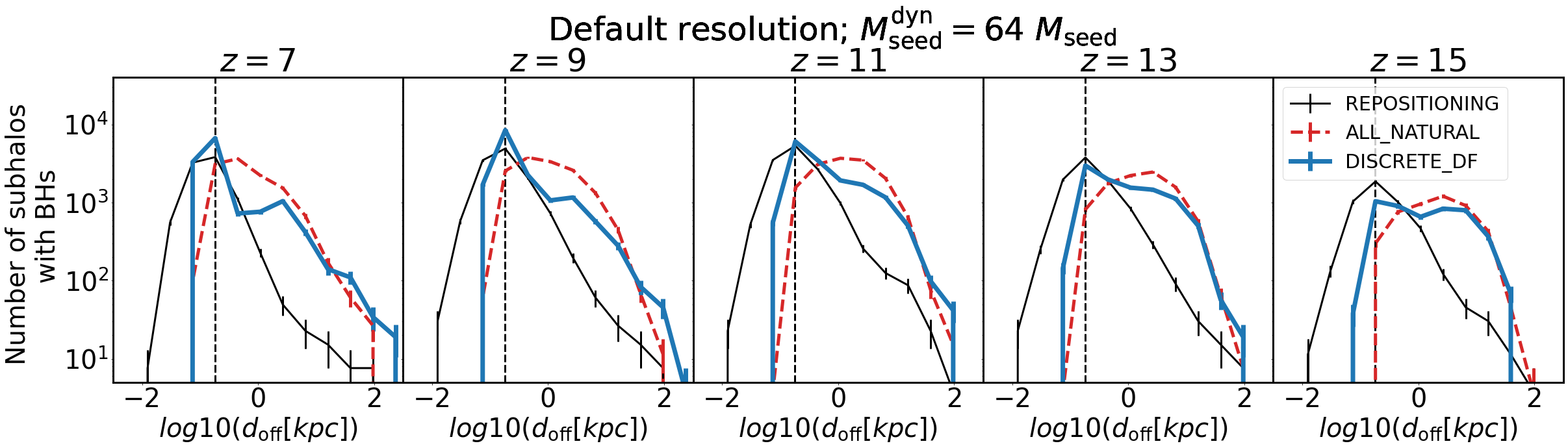}
\includegraphics[width= 18cm]{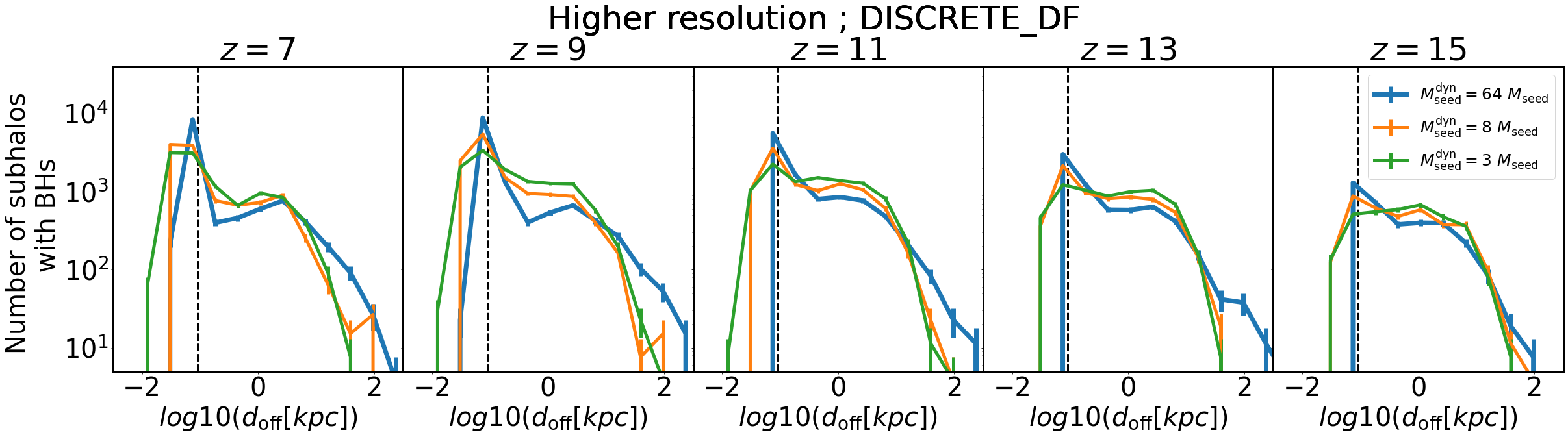}
\caption{\textit{Spatial distributions of BHs inside subhalos:} Distributions of BH offset distances~($d_{\rm off}$) defined as the distance between the most massive BH in a subhalo and potential minima of the subhalo. Note that the contributions from BHs are removed in the potential minima calculation. In the top row, we compare the default resolution \texttt{REPOSITIONING}, \texttt{DISCRETE_DF} and \texttt{ALL_NATURAL} boxes with fixed $\seedmassdynamical=64~\seedmass$. In the bottom row, we have three higher resolution \texttt{DISCRETE_DF} boxes with different values for $\seedmassdynamical$. The subgrid DF does not have a significant impact on the distributions at the highest redshifts~($z\gtrsim15$). However, as time evolves, it becomes increasingly effective and crucial for localizing the BHs at the subhalo centers. By $z=7$, the peaks of the $d_{\rm off}$ distributions for the \texttt{DF_DISCRETE} boxes are close to the gravitational softening lengths, and substantially steeper than the \texttt{ALL_NATURAL} boxes.}
\label{spatial_distributions}
\end{figure*}

Figure \ref{spatial_distributions} shows the distribution of positional offset distances of BHs inside subhalos. We define the offset distance~($d_{\rm off}$) as the distance between the subhalo potential minimum and the most massive BH hosted by it. Since the BH(s) could \textit{alter} the location of the minimum potential toward themselves, in our $d_{\rm off}$ calculation, the contributions of all the BHs within the subhalo are removed while determining the subhalo potential minima. The \texttt{REPOSITIONING} model is naturally the best at strongly localizing the BHs at the subhalo center. Specifically, it produces the most strongly peaked distribution at $d_{\rm off}$ close to the gravitational softening length~(black lines in the top row). Note here that if we include the BH contribution in the potential minimum calculation, the vast majority of the BHs would be \textit{exactly} at the potential minima~($d_{\rm off}=0$) under the repositioning scheme~(by construction). However, we still see a small minority of BHs that are located at distances $\gtrsim 10~\rm kpc$ from the subhalo center. Under the repositioning scheme, these highly offset BHs are possible only in merging systems. 

When repositioning is removed, we see a much larger population of wandering BHs that are offset from the subhalo potential minima. At distances of $\sim1-10~\rm kpc$, the number of BHs increases by factor of $\sim10$ for both \texttt{ALL_NATURAL} and \texttt{DISCRETE_DF} boxes. For the \texttt{ALL_NATURAL} box, the offset distances of BHs tend to peak around $\sim0.5-1~\rm kpc$~(red dashed lines in the top row of Figure \ref{spatial_distributions}), significantly larger than the softening length of the simulations. 

The impact of adding subgrid DF can be readily seen when comparing the offset distances produced by the \texttt{ALL_NATURAL} boxes to those of the \texttt{DISCRETE_DF} boxes~(blue solid vs red dashed lines in the top row of Figure \ref{spatial_distributions}). At $z\sim15$, the \texttt{DISCRETE_DF} simulations do not produce significantly smaller offsets than the \texttt{ALL_NATURAL} boxes, suggesting that the impact of dynamical friction is yet to become significant at these highest redshifts. However, at relatively later redshifts, offsets produced by the \texttt{DISCRETE_DF} boxes become substantially smaller than the \texttt{ALL_NATURAL} boxes. In fact, by $z\sim7$, the $d_{\rm off}$ distributions are much more strongly peaked at the gravitational softening length of the simulations compared to the \texttt{ALL_NATURAL} boxes. The $d_{\rm off}$ distributions peak at the softening length even when we reduce the dynamical seed mass from $64~\seedmass$ to $3~\seedmass$ in the higher resolution boxes~(bottom row of Figure \ref{spatial_distributions}). However, for smaller dynamical seed masses, the peaks of the $d_{\rm off}$ distributions become significantly shallower due to weaker dynamical friction, leading to higher abundances of offset BHs. 


\subsubsection{Velocity distributions of BHs inside subhalos}

\begin{figure*}
\includegraphics[width= 18cm]{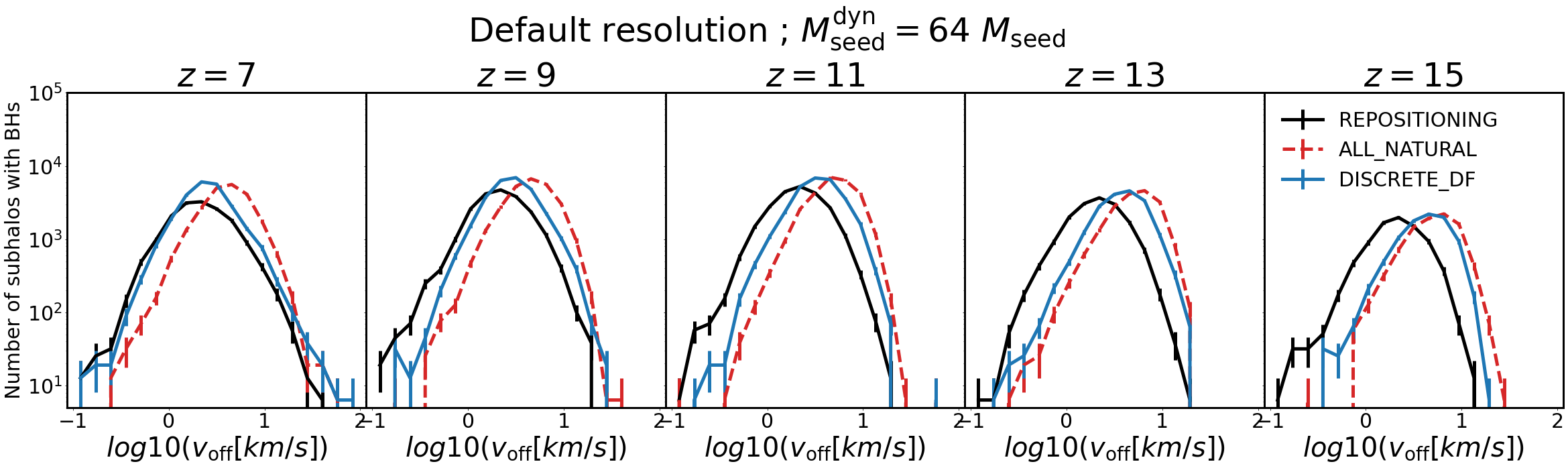}
\includegraphics[width= 18cm]{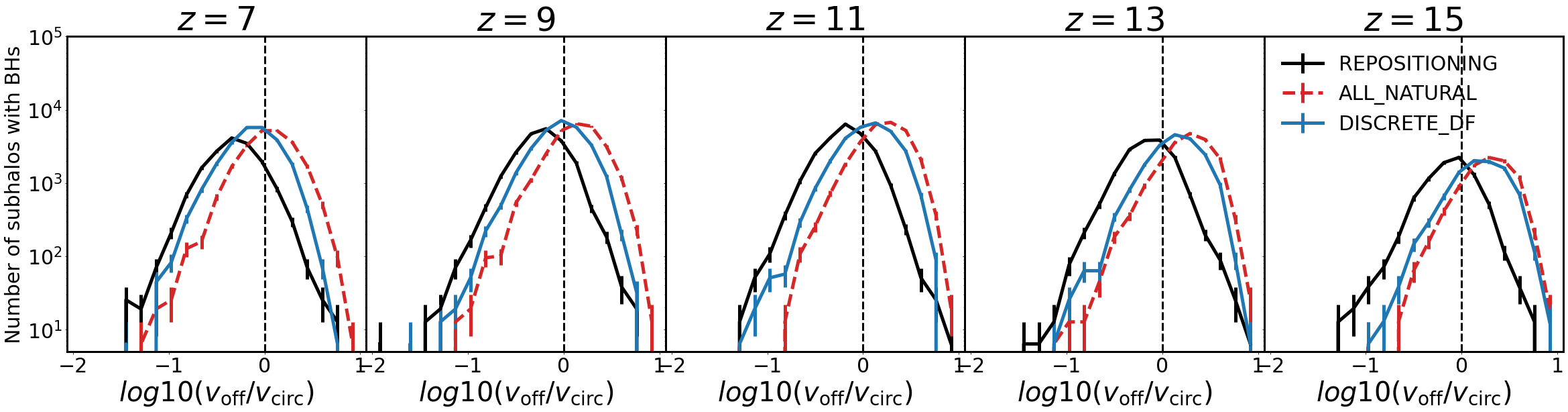}
\includegraphics[width= 18cm]{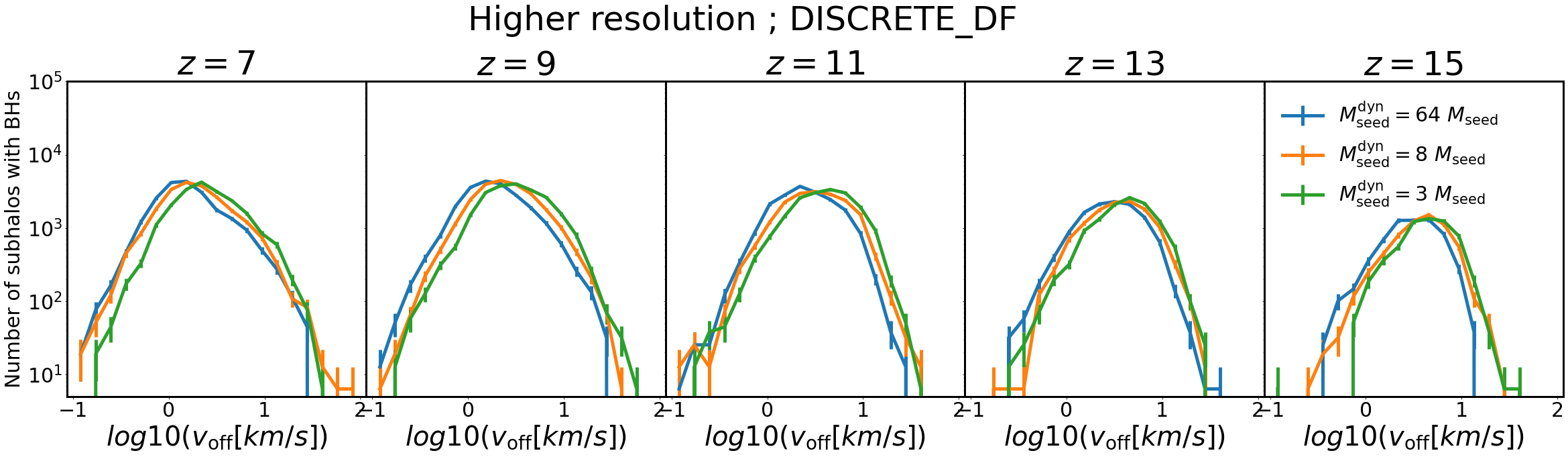}
\caption{\textit{Velocity distributions of BHs inside subhalos:} Distributions of BH velocities~(most massive BH in the subhalo) relative to the center of mass~(of all DM particles) velocities of their host subhalos. The 1st and 2nd rows show the default resolution boxes, wherein the latter is shown in units of the subhalo circular velocity~($v_{\rm circ}$). The 3rd row shows the higher resolution boxes. At the highest redshifts~($z\sim15$), the BH offset velocities predicted by \texttt{DISCRETE_DF} are similar to the \texttt{ALL_NATURAL} boxes. However, by $z\sim7$, \texttt{DISCRETE_DF} predictions are closer to the \texttt{REPOSITIONING} boxes. This again shows that the subgrid DF is more effective at relatively later times.}

\label{velocity_distributions}
\end{figure*}
Figure \ref{velocity_distributions} shows the velocity distributions of BHs hosted by subhalos. Specifically, we compute~(for each subhalo) the relative velocity~(magnitude) of the most massive BH with respect to the average velocity of all DM particles inside the subhalo. We hereafter refer to these velocities as ``offset velocities". 

Let us first focus on the top and middle row which compare the BH offset velocities arising from different dynamics treatments.  For the \texttt{REPOSITIONING} simulations~(black lines), we note that the velocity of the BH at each time-step is essentially set to be the velocity of the minimum potential particle within its immediate vicinity. Not surprisingly, the BHs in the \texttt{ALL_NATURAL} boxes produce substantially larger offset velocities compared to the \texttt{REPOSITIONING} boxes. In the \texttt{DISCRETE_DF} boxes, the offset velocity distributions are similar to that of the \texttt{ALL_NATURAL} boxes at $z=15$. This is similar to what we found for the offset distances, and further suggests that the impact of dynamical friction is yet to become significant at these earliest times. But by $z\sim7$, the offset distributions for the \texttt{DISCRETE_DF} boxes are much closer to the \texttt{REPOSITIONING} model. The middle row specifically shows these offset velcoties in the units of the subhalo circular velocity defined as $v_{\rm circ}\equiv\sqrt{G~M_{\rm sh}/R_{\rm 200}}$ where $M_{\rm sh}$ and $R_{\rm 200}$ is the mass and radius of the host subhalos of the BHs. In the \texttt{ALL_NATURAL} boxes, the peak of the offset velocities are higher than the circular velocities for the majority of BHs at all times. For the \texttt{DISCRETE_DF} boxes, most BHs have velocities less than the circular velocity by $z\sim7$. The bottom row of Figure \ref{velocity_distributions} shows the offset velocity distributions for the three different $\seedmassdynamical$ values in the higher resolution \texttt{DISCRETE_DF} boxes. At all redshifts, the offset velocities tend to be higher for lower values of $\seedmassdynamical$ due to a decrease in the strength of the dynamical friction force.

The key takeaway from these subsections is that subgrid DF plays a key role in effectively sinking a significant majority of BHs to the subhalo centers. However, its impact on the BH populations takes some time to manifest. At $z\sim15$, not enough time has passed for the DF force to have had a substantial impact on BH dynamics. But by $z\sim7$, the impact of the DF force can be readily seen within the BH populations. Finally, the impact of DF is naturally stronger if we assume a higher value for the dynamical seed mass. In the next subsections, we shall look at the implications of these results on the time-scales and rates of BH-BH mergers.

\subsection{Merging BH pairs}
\label{Merging BH binaries}
\begin{figure*}
\centering
\includegraphics[width=15 cm]{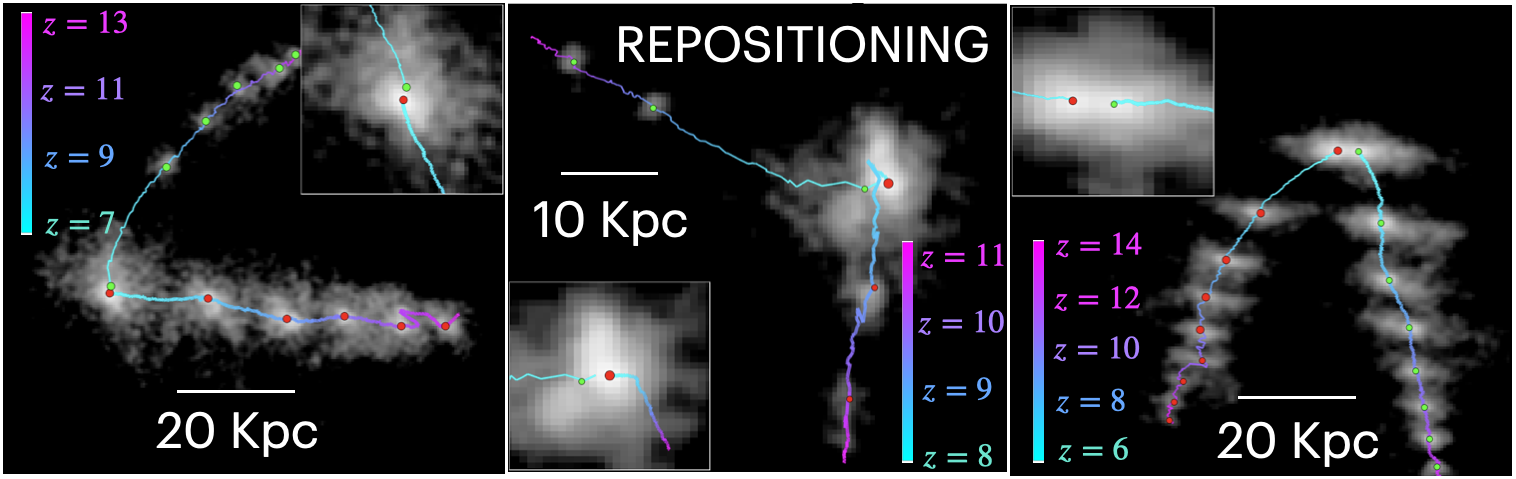}
\includegraphics[width=15 cm]{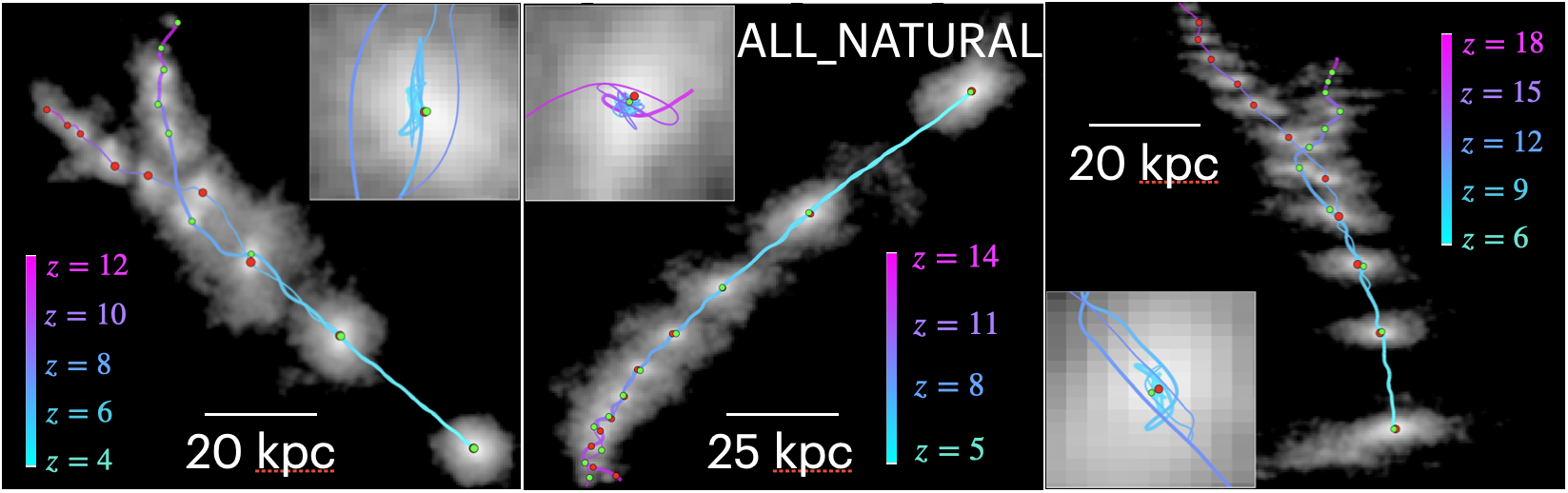}
\includegraphics[width=15 cm]{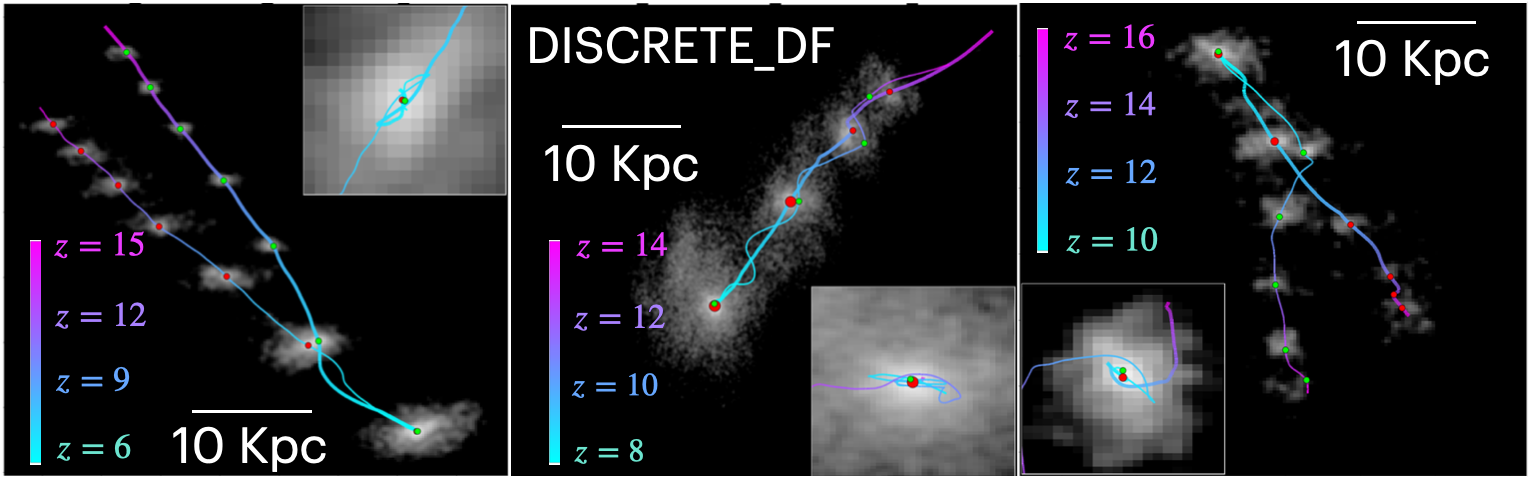}

\caption{Illustrative examples of mergers of BHs and their host subhalos under different dynamics treatments~(using \texttt{MERGE_SOFTENING_BOUND}). In each of the larger panels, the solid lines show the trajectories of the merging BHs wherein the color gradient shows the time evolution. The red and green circles mark the positions of the primary and secondary BHs at different redshift snapshots. At those select snapshots, we also show the DM distributions of the host subhalos of these BHs as grey histograms. The top three panels show examples from the \texttt{REPOSITIONING} boxes, wherein the BHs almost instantaneously merge soon after their host galaxies merge and the BH inspiral is not captured. The middle and lower panels show examples from the \texttt{ALL_NATURAL} boxes and the \texttt{DISCRETE_DF} boxes respectively, that do capture the BH inspiral in the post galaxy merger phase. For each of the larger panels, we have inset panels that show the BH trajectories in the reference frame of the center of mass of the primary and secondary BHs. In the inset panels, the grey histogram shows the DM distribution of the host subhalo in the snapshot closest to the BH merger. In the \texttt{DISCRETE_DF} boxes, the BHs undergo fewer orbits at close separations prior to their eventual merger, compared to the \texttt{ALL_NATURAL} boxes.}
\label{merger_examples}
\end{figure*}

\begin{figure*}
\includegraphics[width=8.5 cm]{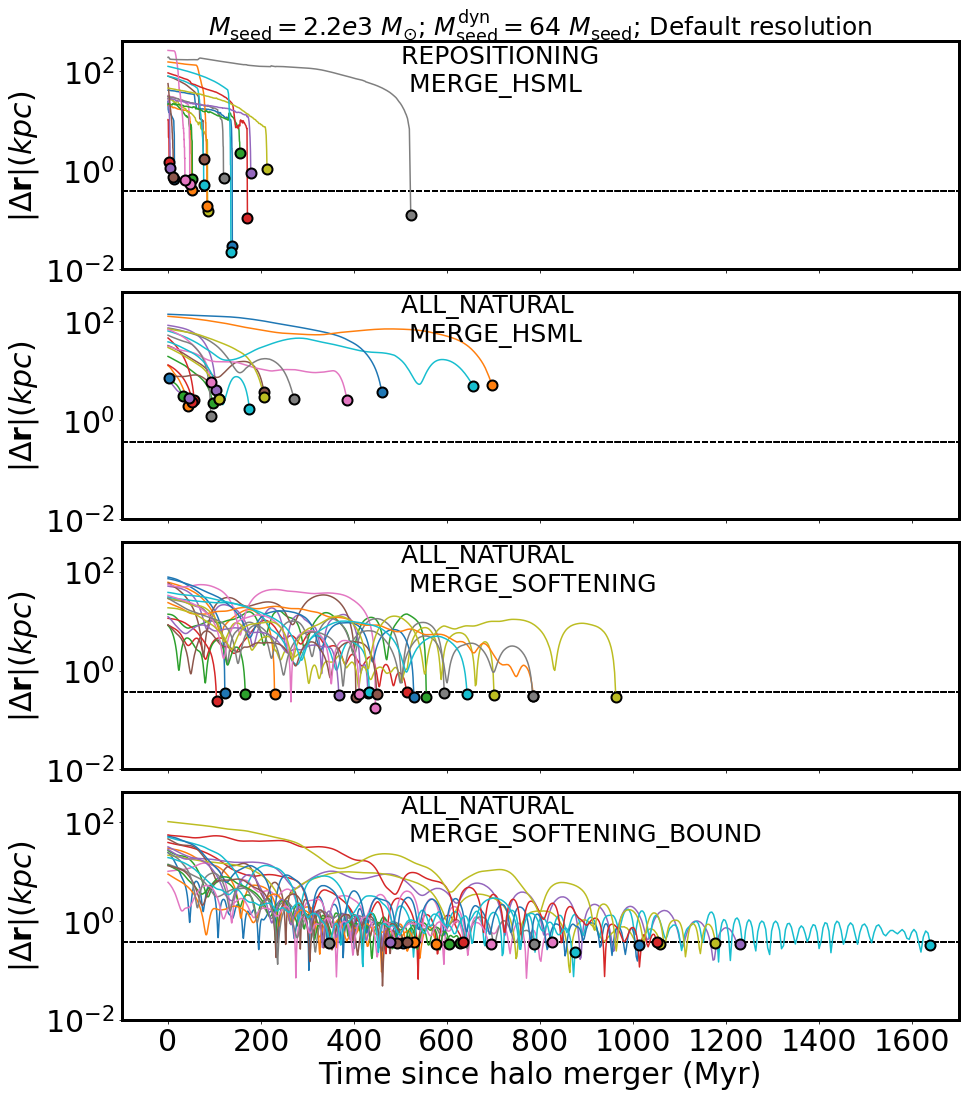}
\includegraphics[width=8.5 cm]{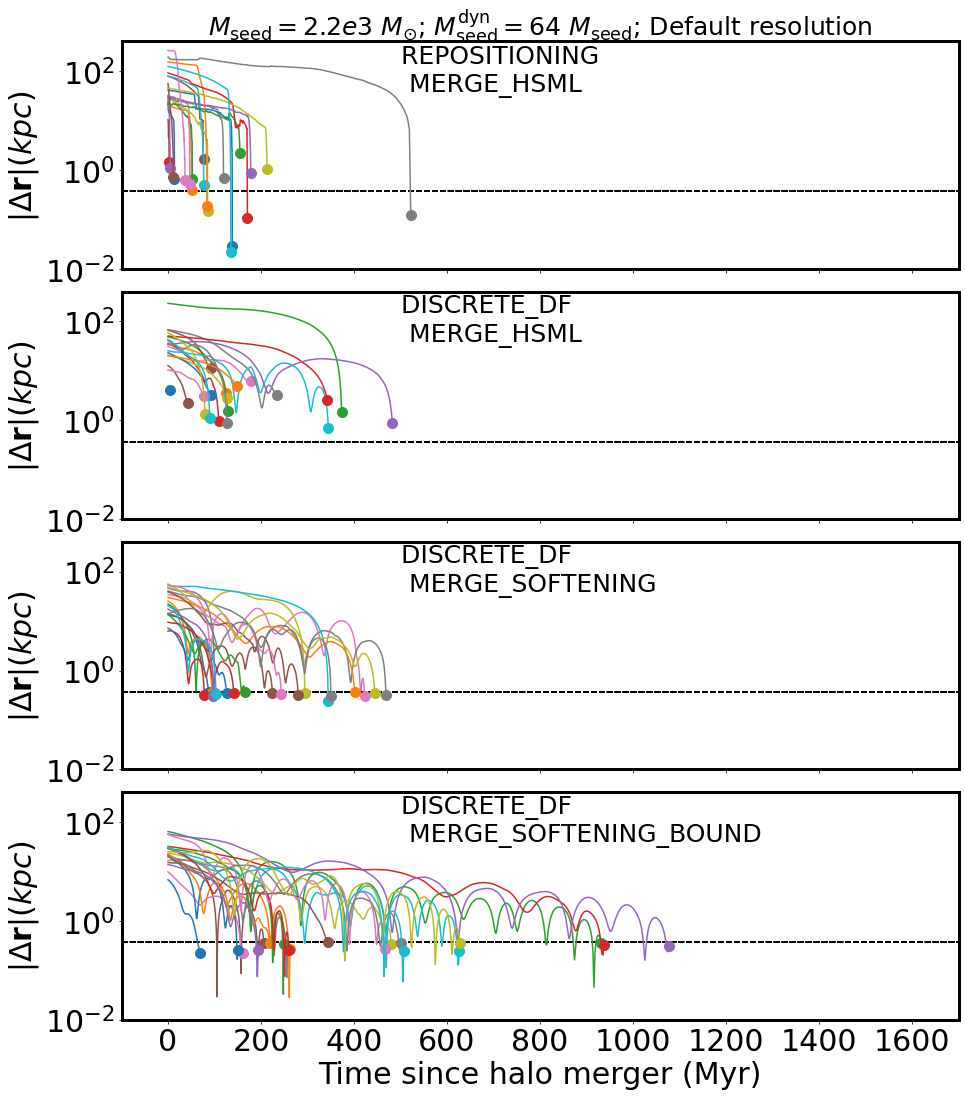}\\

\caption{Evolution of the relative distances of 20 merging BH pairs since their host halos merged, for different BH dynamics treatments as well as the criteria used for merging BHs. The top row shows simulations that use \texttt{REPOSITIONING} and merge BHs within $R_{\rm hsml}$. For the remaining rows, the left panels use \texttt{ALL_NATURAL} and the right panels use \texttt{DF_DISCRETE}. The horizontal dashed line is the minimum separation distance for merging BHs~($2\times$ softening length) Similar to the previous figure, we find that when gravitational boundedness is required for the merging BHs~(\texttt{MERGE_SOFTENING_BOUND}), they merge within fewer orbits in the \texttt{DISCRETE_DF} boxes compared to the \texttt{ALL_NATURAL} boxes.}



\label{time_evolution}
\end{figure*}
Having looked at the dynamics of the full BH populations in our simulations, we now specifically focus on the dynamics of merging BH pairs. We begin by showing visualizations of a few examples of BH mergers and their host halo mergers under different dynamics treatments in Figure \ref{merger_examples}. We show the primary and secondary BHs in the rest frame of the comoving box in the larger panels, and in the frame of the center of mass of the two BHs in the inset panels. For the \texttt{REPOSITIONING} boxes~(top panels), we can clearly see that the trajectories of the BH mergers are not smooth as the BHs get ``teleported" to the nearest potential minima at every time-step. Consequentially, when the host halos merge and no longer have two distinct potential minima, the BHs also merge instantaneously without undergoing any further inspiral. However, in the \texttt{ALL_NATURAL}~(middle) and the \texttt{DISCRETE_DF}~(bottom) boxes, we can readily capture the BH binary inspiral~(up to the spatial resolution) during the post-halo merger phase. Notably, we find in these examples that even when both the \texttt{ALL_NATURAL} and the \texttt{DISCRETE_DF} boxes are able to keep the BHs close to the halo centers, the inspiral process is substantially longer for the \texttt{ALL_NATURAL} boxes. This hints that adding the subgrid DF has a substantial impact in ensuring that the BHs lose enough energy to become gravitationally bound and merge; we shall assess this further for the full BH populations and quantify its impact in the next few subsections.

\subsubsection{Impact of dynamics models and merging criteria}
\label{Impact of dynamics models and merging criteria}
 \begin{figure*}
\centering

\includegraphics[width=18 cm]{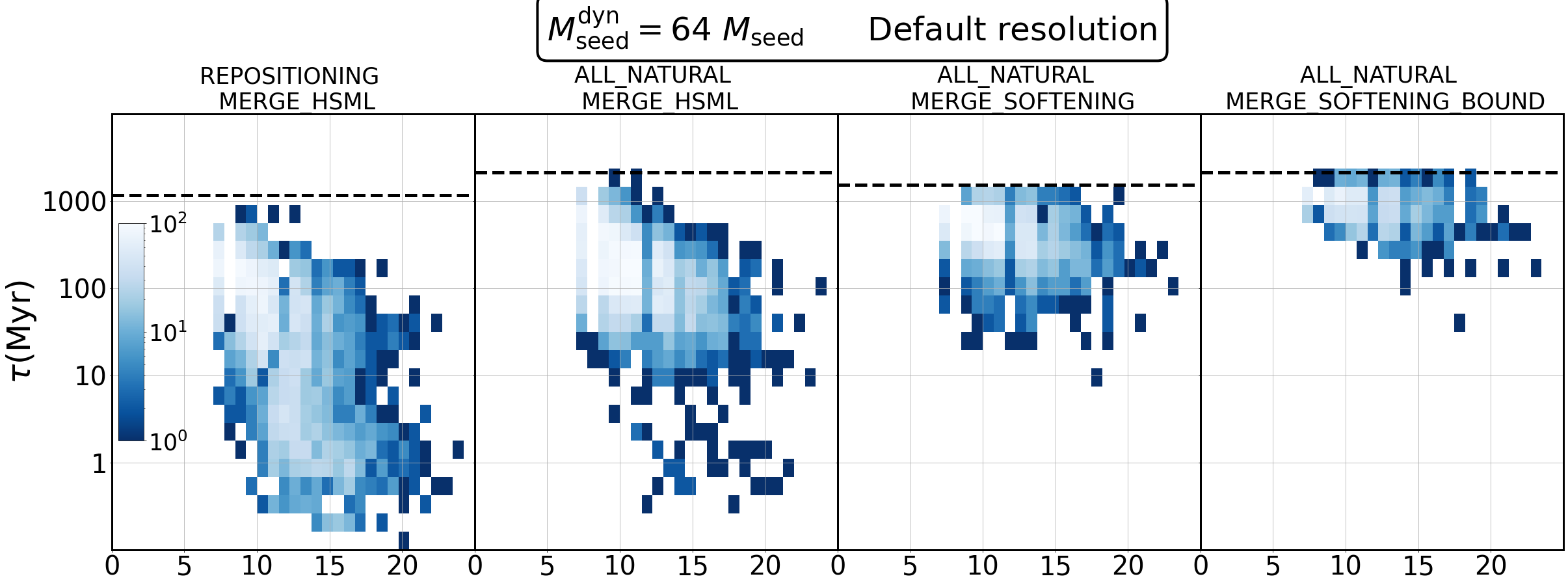}
\includegraphics[width=18 cm]{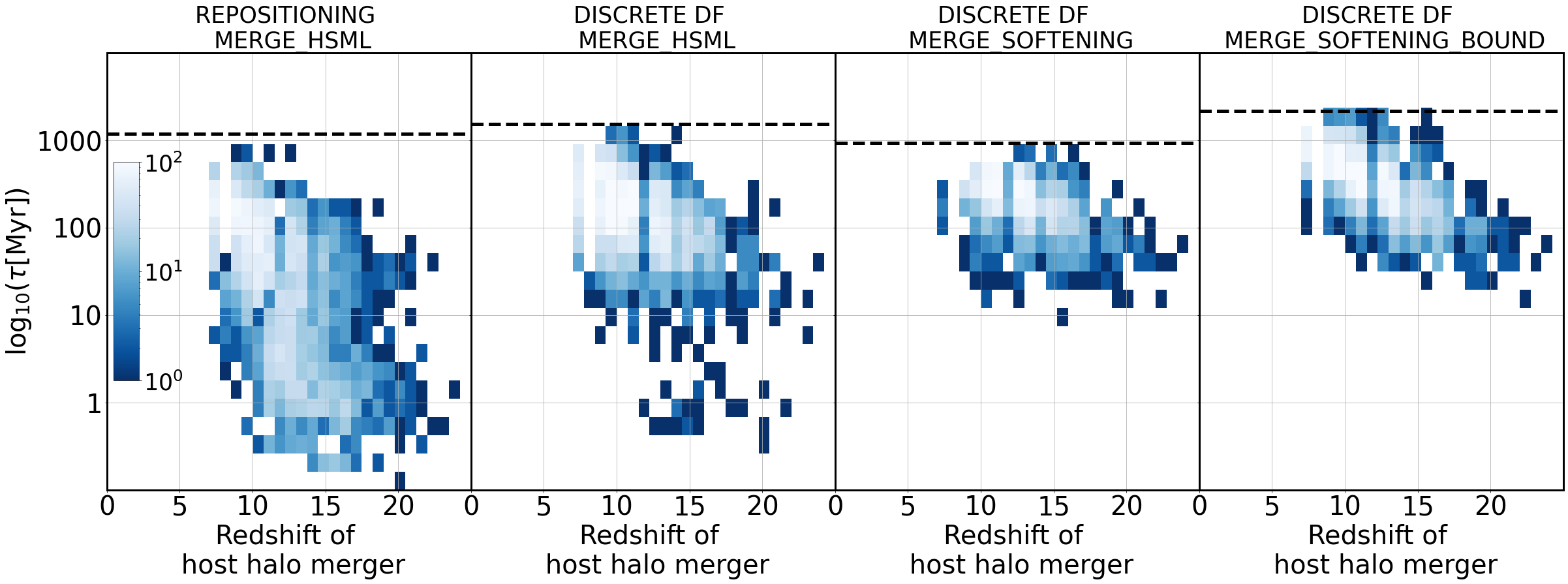}

\caption{Merger time-scales~(time between BH mergers and host halo mergers) of the BH pairs in simulations with different treatments for BH dynamics as well as different merging criteria. 
The left panels show simulations that use \texttt{REPOSITIONING}. For the remaining panels wherein BHs are not repositioned, the top and bottom rows correspond to \texttt{ALL_NATURAL} and \texttt{DISCRETE_DF} simulations respectively. As we go from left to right, the merging criteria becomes more strict. In the first two columns, we use \texttt{MERGE_HSML} wherein BHs are merged within $R_{\rm hsml}\sim2-15~\mathrm{\rm kpc}$. In the 3rd column, we use \texttt{MERGE_SOFTENING} wherein the merging BHs need to be within 2 times the gravitational softening length i.e $\sim0.36~\mathrm{\rm kpc}$. Finally, in the 4th column, we use \texttt{MERGE_SOFTENING_BOUND} that also applies the requirement of gravitational boundedness for the mergers. The dashed lines correspond to the final redshift to which each simulation was run; therefore we do not record mergers with time-scales higher than that. Using the repositioning scheme leads to a substantial number of mergers occurring in $\lesssim10~\rm Myr$ after the BH merger. For the \texttt{DISCRETE_DF} simulations that use \texttt{MERGE_SOFTENING_BOUND}, merger time scales typically range from $\sim50-1000~\mathrm{Myr}$.}

\label{merger_time_scales}
\end{figure*}

We shall now focus on how the different dynamics treatments and merging criteria impact the BH merger time-scales. In Figure \ref{time_evolution}, we show the time evolution of the relative separations of inspiraling BH pairs in the post-galaxy merger phase for a randomly selected set of 20 merger events. Subsequently, we plot the time-scales of all the merger events in Figure \ref{merger_time_scales}. We define the merger timescale $\tau$ as the time elapsed between BH-BH merger and the merger of the host halo. Note that the host halo merger time can only be computed to the precision of the snapshot resolution, as halo outputs are only available during snapshots. On the other hand, the BH evolution and mergers are tracked at the~(much higher) time resolution of the simulation. Our methodology for estimating the halo merger time is described in Appendix \ref{halo_merger_estimate}.  

\begin{figure*}
\centering
\includegraphics[width=16 cm]{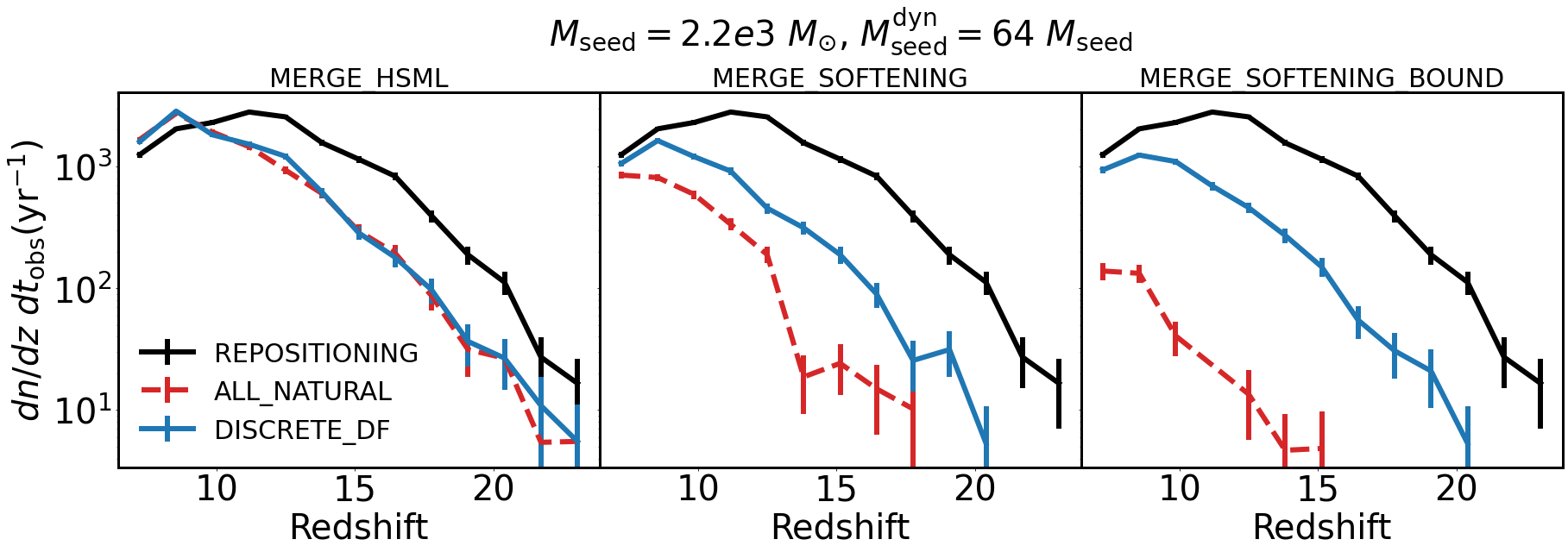}
\caption{Predicted merger rates of $2.2\times10^3~M_{\odot}$ seeds with $\seedmassdynamical = 64~\seedmass$ in simulations with different treatments of BH dynamics and merging criteria. The left, middle and right panels correspond to \texttt{MERGE_HSML}, \texttt{MERGE_SOFTENING} and \texttt{MERGE_SOFTENING_BOUND}. In each panel, we compare the predictions from \texttt{REPOSITIONING}, \texttt{ALL_NATURAL} and \texttt{DF_DISCRETE} boxes. The subgrid DF plays the most crucial role in merging BHs when we require gravitational boundedness~(\texttt{MERGE_SOFTENING_BOUND}). }
\label{merger_rates_different_models}
\end{figure*}

We now look at Figures \ref{time_evolution}, \ref{merger_time_scales} and \ref{merger_rates_different_models} together to understand the impact of the dynamics treatment and merging criteria on the merger time scales and merger rates. When repositioning is used~(Figure \ref{time_evolution} 1st row and Figure \ref{merger_time_scales} 1st column), BHs typically merge at separations of several kpc, depending on their neighbor search radii~($R_{\rm hsml}$) over which the local potential minimum is determined~(top row of Figure \ref{time_evolution}). The resulting merger time scales can range from several hundred Myrs all the way down to less than a Myr~(1st row of Figure \ref{merger_time_scales}). When repositioning is removed,  we see a drastic reduction in merger events with time scales $\lesssim10~ \rm Myr$~(1st vs 2nd row of Figure \ref{merger_time_scales}). In fact, when the merging distance is reduced from $R_{\rm hsml}$ to $2~\epsilon_g$, we find that merger time-scales less than $10~\mathrm{Myr}$ are no longer possible for both \texttt{ALL_NATURAL} and the \texttt{DISCRETE_DF} boxes~(Figure \ref{merger_time_scales} 3rd and 4th columns).  

\begin{figure}
\hspace{-1cm}
\includegraphics[width=9 cm]{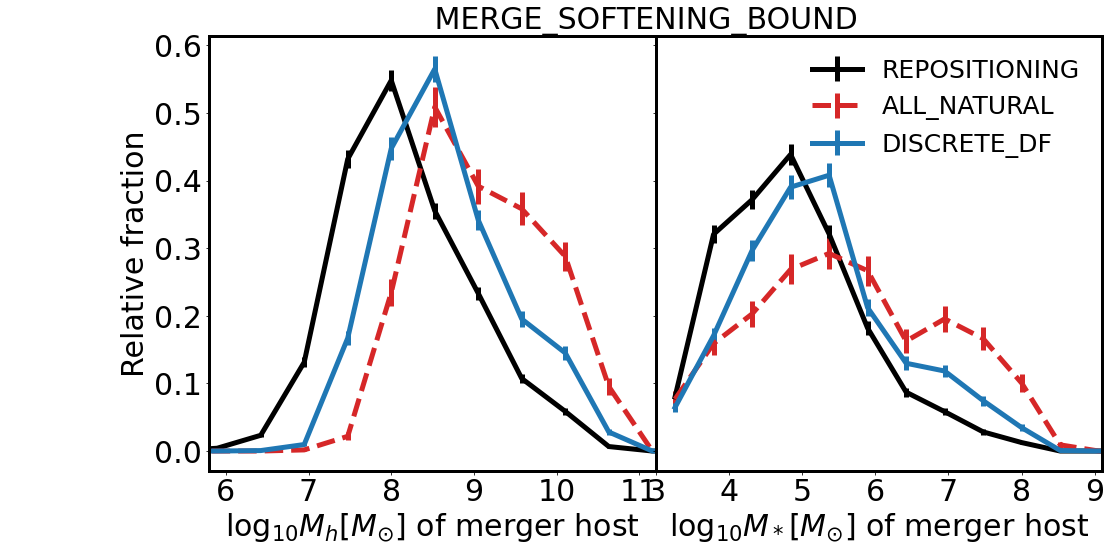}
\caption{Distributions of DM masses~(left) and stellar masses~(right) of the host halos of the BH merger events for the \texttt{REPOSITIONING}, \texttt{ALL_NATURAL} and \texttt{DISCRETE_DF} boxes~(with $\seedmassdynamical=64~\seedmass$) using \texttt{MERGE_SOFTENING_BOUND}. For the \texttt{DISCRETE_DF} boxes, most mergers occur in $\sim10^8~M_{\odot}$ halos. The \texttt{REPOSITIONING} boxes merge BHs in lower mass halos~(by factors of $\sim3$), whereas the \texttt{ALL_NATURAL} boxes merge BHs in relatively higher mass halos~(also, by factors of $\sim3$) compared to \texttt{DISCRETE_DF}.  }
\label{merger_hosts}
\end{figure}

For the boxes without repositioning, let us now focus on the impact of adding subgrid DF by comparing the left vs. right columns of Figure \ref{time_evolution} and upper vs. lower panels of Figure \ref{merger_time_scales}. When the merging distances are set to be $R_{\rm hsml}$~(\texttt{MERGE_HSML}), which is $\sim$ several kpc, the addition of the subgrid DF does not strongly impact the merger time-scales. This can be much more readily seen in the merger rates~(left panel of Figure \ref{merger_rates_different_models}), wherein the predictions of the \texttt{DISCRETE_DF} boxes are only slightly higher than the \texttt{ALL_NATURAL} boxes. Essentially, this means that the subgrid DF is not consequential to assembly of $\sim \rm kpc$ scale pairs in our simulations. This is not surprising given that these pairs assemble in $\sim10^8~M_{\odot}$ halos~(see left panels of Figure \ref{merger_hosts}), with virial radii~($\sim8~\rm kpc$) that are not much larger than the merging separations. In other words, the halo mergers alone are enough to bring the binary separations close to a few kpc scales, without the need for subgrid DF. 

Notably, Figure \ref{merger_hosts} also shows that the typical masses of the host halos for the BH mergers in the \texttt{REPOSITIONING} boxes are lower than the \texttt{DISCRETE_DF} boxes by factor of $\sim3$. This is expected due to the shorter BH merging time scales with BH repositioning, leaving less time for the halos to grow before the BHs merge. In the same vein, the typical host halos that merge BHs in the \texttt{ALL_NATURAL} boxes are a factor of $\sim3$ larger than the \texttt{DISCRETE_DF} boxes.   

When the merger separation is set to $2~\epsilon_g = 0.36~\rm kpc$~(\texttt{MERGE_SOFTENING}), which is much smaller than the virial radii of the merging halos, BH pairs now have to undergo a substantial amount of orbital decay to merge in the simulations. In this case, the \texttt{ALL_NATURAL} boxes produce noticeably higher merger time-scales than the \texttt{DISCRETE_DF} boxes. Concurrently, the merger rates for the \texttt{ALL_NATURAL} boxes are a factor $\sim6$ smaller than the \texttt{DISCRETE_DF} boxes~(middle panel of Figure \ref{merger_rates_different_models}).  

Finally, when the requirement of gravitational boundedness is also applied to the merging criteria~(\texttt{MERGE_SOFTENING_BOUND}), we find that the role of subgrid DF becomes much more crucial compared to when we merge BHs solely based on distance. In the \texttt{ALL_NATURAL} boxes, we find that many of the BHs that would have merged at relatively earlier times using \texttt{MERGE_SOFTENING} undergo many more rounds of inspiral before losing enough energy to merge using \texttt{MERGE_SOFTENING_BOUND}~(see Figure \ref{merger_examples} middle panels and Figure \ref{time_evolution} bottom left panel). When we apply subgrid DF, it leads to a substantial reduction in the number of inspirals prior to merger, as the BHs are able to lose energy much more efficiently. In the rightmost panels of Figure \ref{merger_time_scales}, we therefore find a  substantial reduction in the merger time-scales within the \texttt{DISCRETE_DF} boxes compared to the \texttt{ALL_NATURAL} boxes when \texttt{MERGE_SOFTENING_BOUND} is applied. Specifically, the \texttt{DISCRETE_DF} boxes produce merger time-scales that span between $\sim50-1000~\rm Myr$, whereas those in the \texttt{ALL_NATURAL} boxes are almost always $\gtrsim200~\rm Myr$. In terms of the merger rates~(rightmost panels of Figure \ref{merger_rates_different_models}), the \texttt{ALL_NATURAL} boxes produce $\sim10$ times fewer mergers compared to the \texttt{DISCRETE_DF} boxes when \texttt{MERGE_SOFTENING_BOUND} is applied.

The key takeaway from the above discussion is that subgrid DF is crucial to ensure that a significant fraction of the BH pairs formed during halo mergers eventually merge with one another. They play a particularly critical role in ensuring that the BH pairs lose enough energy to eventually become gravitationally bound.

\begin{figure*}
\centering
\includegraphics[width=5.5 cm]{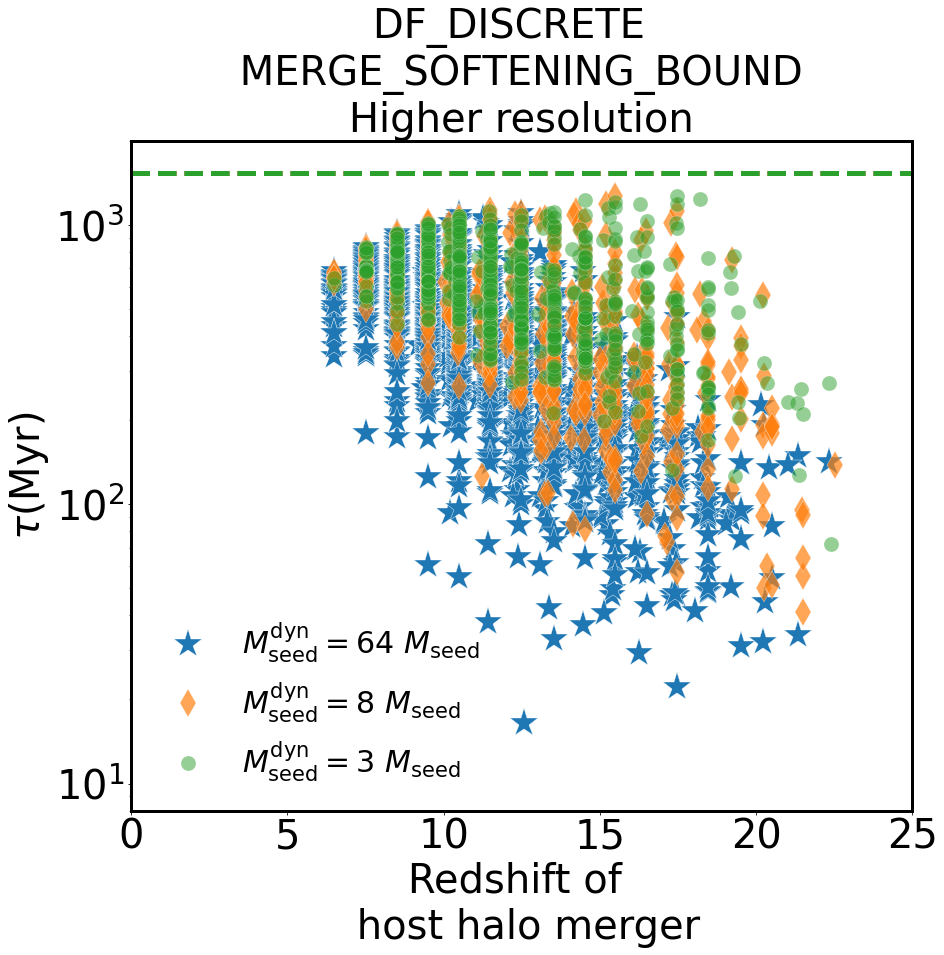}
\includegraphics[width=5.5 cm]{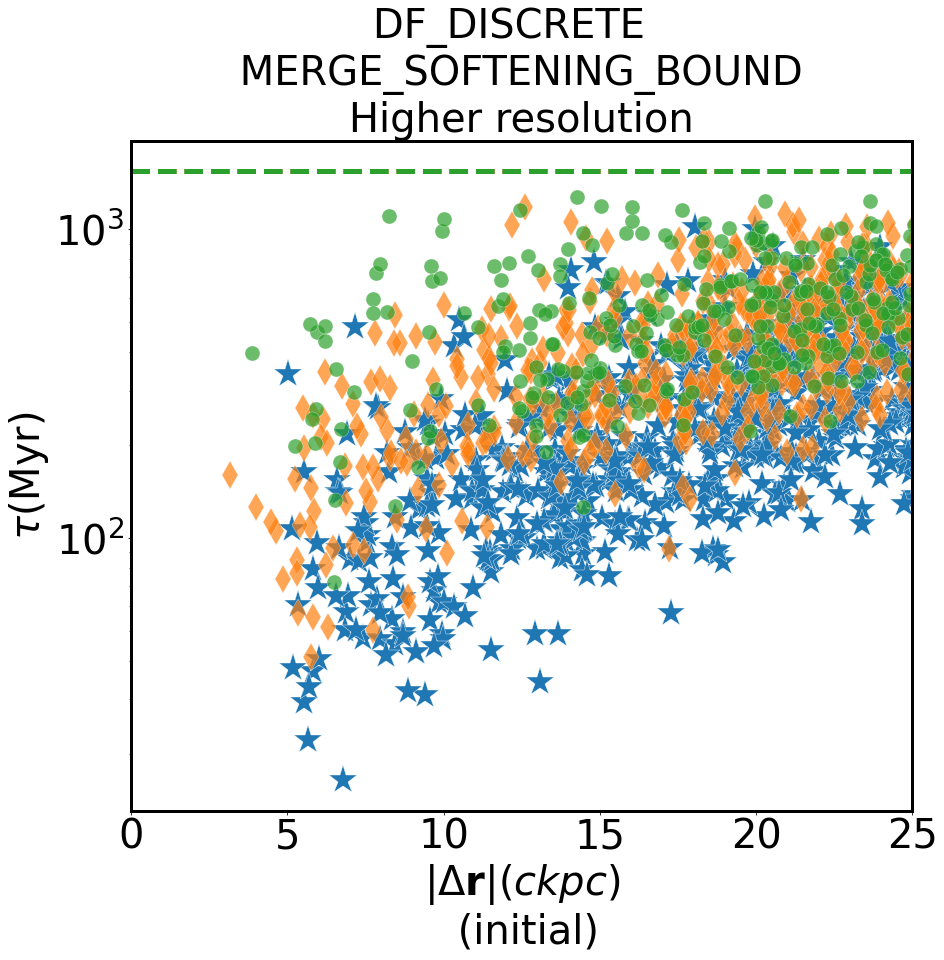}
\includegraphics[width=5.5 cm]{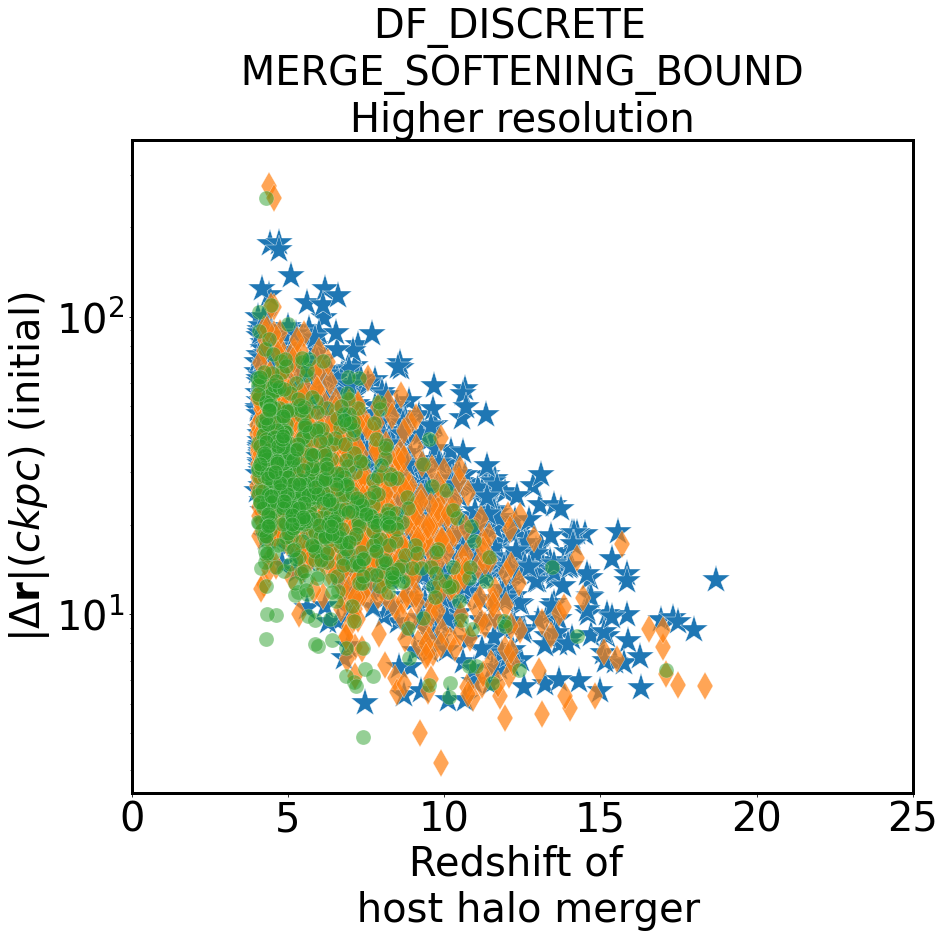}

\caption{Left and middle panels show BH merger time-scales plotted against the redshift of the host halo merger and the initial binary separation~(at the time of halo merger) respectively. Blue, orange and green points show $\seedmassdynamical=3,8~\&~64~\seedmass$ respectively. Green dashed horizontal lines correspond to the simulation end time. The right panel plots the initial binary separation as a function of host halo merger redshift. Altogether, the plots demonstrate that for the high redshift halo mergers, the initial separations of the BH pairs are smaller~(even in comoving units). As a result, higher redshift halo mergers lead to BH mergers at relatively smaller time-scales.}
\label{merger_time_scale_redshift_dependence}
\end{figure*}

\subsubsection{Redshift dependence of the merger time-scales}
Figure \ref{merger_time_scales} reveals another noteworthy feature for both \texttt{REPOSITIONING} and \texttt{DISCRETE_DF} boxes: the merger time scales tend to be shorter for host halos that merged at higher redshifts. This seems a bit surprising given the results from Section \ref{Dynamics of BHs inside their host subhalos} which showed that subgrid DF is generally less effective at higher redshifts in localizing the BHs to the halo centers. The left panel in Figure \ref{merger_time_scale_redshift_dependence} shows that this redshift dependence is true regardless of our choice for the dynamical seed mass. Note also that this redshift dependence is most apparent for halo mergers that occurred well before the simulation end time~(indicated by dashed vertical lines in Figure~\ref{merger_time_scale_redshift_dependence}). This is simply because mergers that would have taken place after the simulation end time are inherently missing, and such late-time mergers are more common for halos that merge at lower redshifts.


We further investigate why the BH merger time scales tend to be smaller for the higher redshift halo mergers by looking at the initial separations~(i.e. at the time of halo merger) of the merging BH pairs. The middle panel of Figure \ref{merger_time_scale_redshift_dependence} unsurprisingly shows that when the initial separations of the BH pairs~(at the time of halo merger) are smaller, the merger time scales are also shorter. But more crucially, the right panel finally shows that these initial separations are smaller when their host halos merged at higher redshifts. This overall means that BH merger time scales at higher redshifts are shorter because their host halo mergers already leave them at smaller separations. Note also that these separations are plotted in co-moving units, thereby making this redshift dependence an additional effect that occurs on top of cosmological expansion. Adding the impact of cosmological expansion would lead to even smaller initial separations~(in proper units) at high redshifts, thereby facilitating faster orbital decay and shorter merger-time scales. In summary, these results indicate that shorter merger time-scales at higher redshifts are due to smaller separations of inspiraling BH pairs during the initial stages of the post-halo-merger phase.

\subsubsection{Impact of dynamical seed mass}

\label{Impact of dynamical seed mass}

\begin{figure*}
\centering
\includegraphics[width=18 cm]{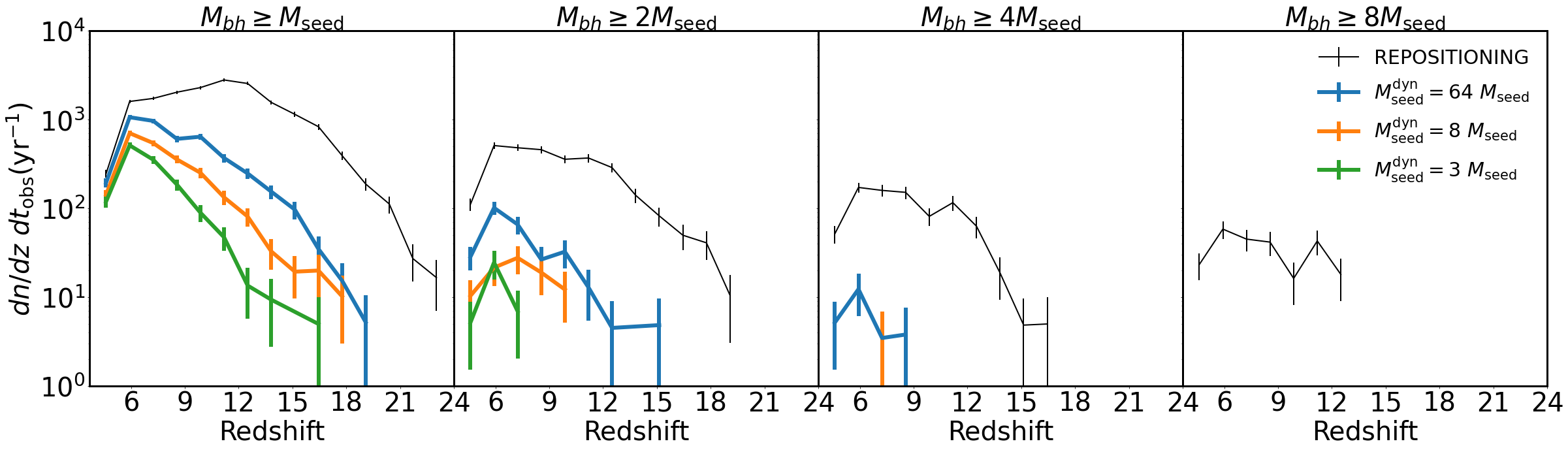}
\includegraphics[width=18 cm]{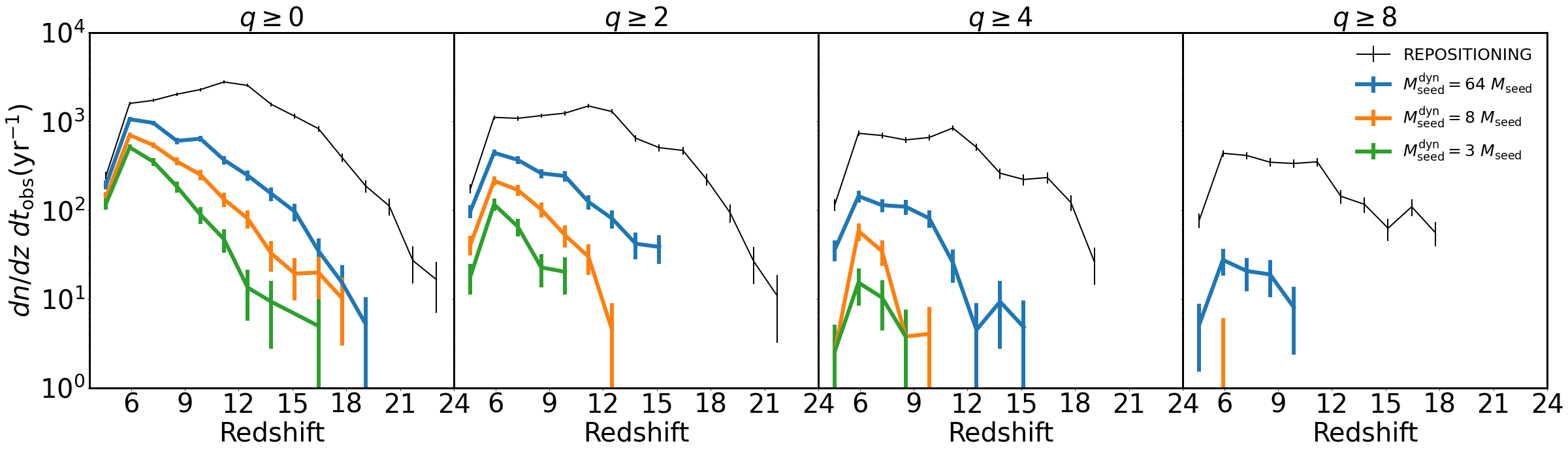}

\caption{Predicted merger rates of $2.2\times10^3~M_{\odot}$ seeds with $\seedmassdynamical \sim 3,8~\&~64~\seedmass$ respectively. The blue, orange and green lines are from the higher resolution \texttt{DISCRETE_DF} boxes that use \texttt{MERGE_SOFTENING_BOUND}. The black line shows the prediction using \texttt{REPOSITIONING}. In the top panels, going left to right increases the masses of both the primary and secondary BHs. The left most panels show the overall merger rates. In the bottom panels, going left to right increases the mass ratios between the primary and the secondary. Compared to repositioning, the overall merger rates at $z\gtrsim5$ are suppressed by factors of $\sim4-10$ depending on the dynamical seed masses.}

\label{merger_rates_different_dynamical_seed_masses}
\end{figure*}

Having discussed how the different dynamics treatments and merging criteria influence the merging BH populations produced by our simulations, we will now take our fiducial setup~(i.e. \texttt{DISCRETE_DF} combined with \texttt{MERGE_SOFTENING_BOUND}) and look at how they are influenced by our assumption of the dynamical seed mass~($\seedmassdynamical$) in the higher resolution boxes. Figure \ref{merger_time_scale_redshift_dependence} shows that the merger time scales tend to increase when $\seedmassdynamical$ is decreased, even if they start at similar separations in the post halo merger phase. This is not at all surprising as the dynamical friction will be weaker for lower values of $\seedmassdynamical$. Figure \ref{merger_rates_different_dynamical_seed_masses} shows the merger rates produced by different values of $\seedmassdynamical$. As we decrease the  $\seedmassdynamical$, we see a suppression in the merger rates due to the increased merger time-scales. 

In general, the strongest suppression is seen at the highest redshifts even though the merger time-scales are smaller at higher $z$; this is simply because a fixed time delay corresponds to a larger redshift interval at high $z$. However, many mergers do happen by our final redshift of $z\sim5$. Therefore, the overall suppression in the $z\gtrsim5$ merger rates due to dynamical delays is much smaller compared to when we exclusively focus on the high redshift~(e.g. $z\gtrsim15$) mergers. 

It is useful to quantify the overall suppression in the merger rates for different dynamical seed masses by comparing against the \texttt{REPOSITIONING} boxes that produce uppermost limits to the merger rate for a given seed model. Let us first focus on the overall merger rates shown in the left-most panels of Figure \ref{merger_rates_different_dynamical_seed_masses}. While the \texttt{REPOSITIONING} boxes produce 4316 mergers at $z\gtrsim5$, the \texttt{DISCRETE_DF} boxes produce 1223, 663 and 403 mergers for $\seedmassdynamical = 64, 8 ~\&~ 3~\seedmass$ respectively. This implies an overall suppression of $\sim4-10$ in the $z\gtrsim5$ merger rates compared to the \texttt{REPOSITIONING} boxes. At the highest redshifts of $z\gtrsim15$,  merger rates are suppressed by factors of $\sim10-100$ compared to the \texttt{REPOSITIONING} boxes. 

The 2nd, 3rd and 4th panels in the top row of Figure \ref{merger_rates_different_dynamical_seed_masses} show the merger rates for higher mass thresholds of $2\seedmass,4\seedmass,~\&~8\seedmass$ for both the primary and secondary masses. Note that since the BH growth is primarily driven via mergers~(as we shall see in Section \ref{Mergers vs accretion driven BH growth}) and the seed mass is assumed to be uniform throughout, $M_{bh} \geq 2\seedmass$ essentially corresponds to second generation mergers. The rates of these mergers are $\sim10-50$ times lower compared to the 1st generation mergers depending on the assumed $\seedmassdynamical$. Higher mass thresholds correspond to further generations of mergers. The rates of $M_{bh} \geq 4\seedmass$ merger events, which include the third generation mergers, are $\gtrsim100$ times smaller than the first generation mergers. For the \texttt{DISCRETE_DF} boxes, the rates of higher generation mergers naturally decline much more steeply compared to the \texttt{REPOSITIONING} boxes, simply because of the smaller number of first generation mergers that are needed to create the remnants for subsequent generations of mergers. For the same reason, our ability to generally probe higher generations of mergers in the \texttt{DISCRETE_DF} boxes is very limited by our small simulation volume. 

Finally, the 2nd, 3rd and 4th panels in the bottom row of Figure \ref{merger_rates_different_dynamical_seed_masses} show the merger rates at higher thresholds of mass ratios~($q$) between the primary and the secondary. For $q>1$, the primary BHs are remnants of previous generations of mergers. Here too, the \texttt{DISCRETE_DF} boxes show a steeper decline in the merger rates at higher mass ratios compared to the \texttt{REPOSITIONING} boxes, simply because of the lower number of first generation mergers. Depending on the dynamical seed masses, the merger rates for $q>2$ events are $\sim2-5$ times smaller than the overall merger rate, and that for the $q>4$ events are $\sim10-100$ times smaller. Above these mass ratios, our simulations do not produce many mergers due to their limited volumes.     
 
\subsubsection{Merger rates for different seed models}
\label{Merger rates for different seed models}

\begin{figure*}
\centering
\includegraphics[width=14 cm]{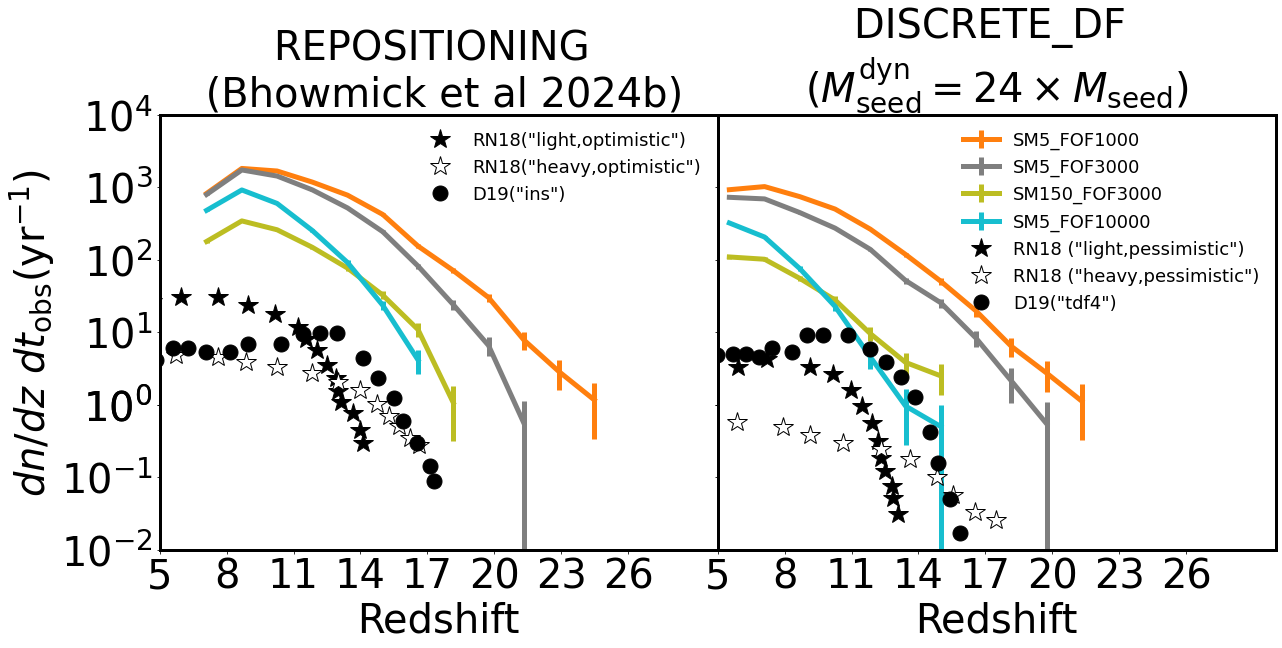}
\caption{Merger rates for all four seed models. The left panel shows the \texttt{REPOSITIONING} model and the right panel shows the \texttt{DISCRETE_DF} model together with the \texttt{MERGE_SOFTENING_BOUND} merger criterion. For the \texttt{DISCRETE_DF} boxes, we apply $\seedmassdynamical = 24~\seedmass$. In the left panel, dashed lines show results from \protect\cite{2024MNRAS.531.4311B} which use $9\rm ~Mpc$~(\texttt{BRAHMA-9}) boxes. For the \texttt{DISCRETE_DF} boxes, the different seed models predict $z\gtrsim5$ merger rates of $100-1000~\rm yr^{-1}$. } 
\label{merger_rates_gas_based_seeding}
\end{figure*}

For the vast majority of this paper, we have used the \texttt{BRAHMA-4.5} boxes and looked at the impact of different dynamics treatments and merging criteria for a fixed seed model of \texttt{SM5_FOF1000}. Having done that, we now choose a fiducial dynamics setup and use the \texttt{BRAHMA-9} boxes to provide BH merger rate predictions for all four seed models introduced in Section \ref{Black hole seeding}. For our fiducial dynamics setup, we used \texttt{DISCRETE_DF} combined with \texttt{MERGE_SOFTENING_BOUND}. We use a dynamical seed mass of $\seedmassdynamical = 24~\seedmass$ which is about $\sim2.3~M_{\rm DM}$~(a BH mass to DM mass ratio that we already tested with $\seedmassdynamical = 3~\seedmass$ in the higher resolution boxes). We do this to close the gap between $\seedmassdynamical$ and $\seedmass$ as much as possible while still using the default resolution boxes. The merger rate predictions are presented in the right panel of Figure \ref{merger_rates_gas_based_seeding}. For comparison, we also show predictions from our previous \texttt{BRAHMA-9} boxes~\citep{2024MNRAS.531.4311B} that used the repositioning scheme~(left panel of Figure \ref{merger_rates_gas_based_seeding}). 

While BH repositioning yields merger rates (upper limits) that peak at $z \sim 9$–$12$, reaching values of $\sim200$–$2000~\rm yr^{-1}$ depending on the seed model, the merger rates from our subgrid DF model continue to rise until $z \sim 5$. At $z \sim 5$, the subgrid DF model predicts merger rates of $\sim100$–$1000~\rm yr^{-1}$, again depending on the seed model. Although the peak merger rates in the subgrid DF model are only lower by a factor of $\sim2$ compared to repositioning, the difference becomes much more pronounced at the highest redshifts. For instance, at $z \gtrsim 14$, merger rates are $\sim50$–$800~\rm yr^{-1}$ with repositioning, but only $\sim1$–$100~\rm yr^{-1}$ with the subgrid DF model. 

Note that even without repositioning, our subgrid DF-based predictions should still be regarded as upper limits—albeit more conservative ones—since we do not track the inspiral of BH pairs below the gravitational softening scale, nor do we include the effects of gravitational recoil kicks to the merger remnants. We discuss these caveats in more detail in the next section. 

It is instructive to compare our predictions against previous work, particularly those that make overall merger rate predictions for ``light" seed models such as the semi-analytic models of \citealt[][(R18)]{2018MNRAS.481.3278R} and \citealt[][(D19)]{2019MNRAS.486.2336D}. In \cite{2024MNRAS.531.4311B}, we ensured an even-handed comparison with the repositioning-based predictions by looking at only those SAM predictions that assumed instantaneous mergers of BHs after their host halos merge~(black points in the left panel of Figure \ref{merger_rates_gas_based_seeding}). For the subgrid DF based predictions of this work, we maintain a similar level of even-handedness by comparing to those predictions that accounted for potential dynamical delays. For example, R18 performs a ``pessimistic" prediction~(black stars in the right panel) where only $10~\%$ of galaxy mergers led to a BH merger. The ``tdf4" model of D19~(black circles in the right panel) computed a merging time scale between the galaxy (along with their BH) mergers and the host halo mergers based on the dynamical friction force encountered by the satellite galaxies~(see Eq. 23 in their paper). Regardless of the dynamics modeling, our predictions are $\sim10-100$ times higher than both these SAMs. This is primarily because the seed models of R18 and D19 are much more restrictive than the \texttt{BRAHMA} simulations. Our \texttt{BRAHMA} boxes allow seed formation at all redshifts within $\sim10^6-10^7~M_{\odot}$ halos depending on the seed model. On the other hand, R18 only allows seed formation between $z\sim15-20$, while D19 restricts seed formation to $\gtrsim10^8~M_{\odot}$ halos.  

Overall, the high rates of mergers in our simulations are an exciting prospect for LISA. Based on our results, we could expect a substantial number of LISA events during its lifetime even if only a small fraction~($\sim1-10~\%$ for example) of them end up coalescing and producing GW waves. 

\begin{figure*}
\centering
\includegraphics[width= 18cm]{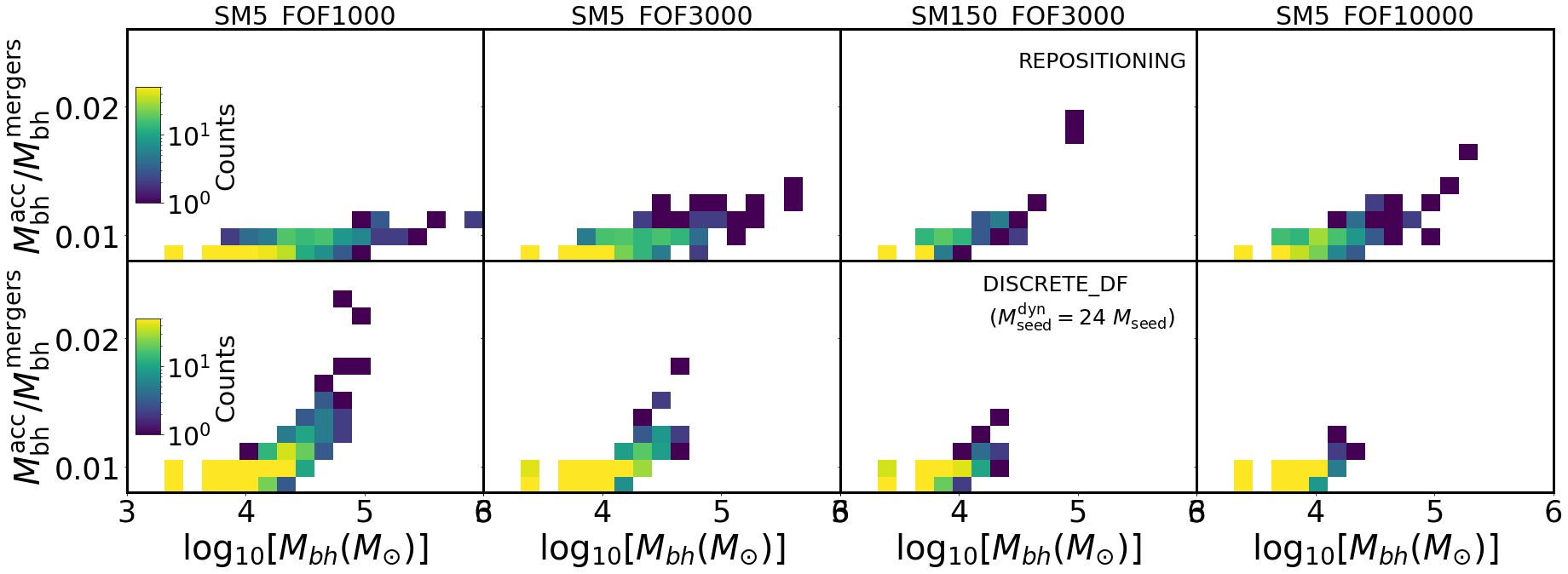}

\caption{For the four seed models in the previous figure, we show the fractional growth in BH mass that occurs due to gas accretion instead of BH-BH mergers. We plot these as 2D histograms as a function of BH mass for BH populations at $z=5$. The initial BH growth is dominated by mergers for both the \texttt{REPOSITIONING} as well as the \texttt{DISCRETE_DF} boxes~(using \texttt{MERGE_SOFTENING_BOUND}). Under subgrid DF, mergers can grow the BHs up to masses ranging between $\sim10^4-10^5~M_{\odot}$ by  $z\sim5$ depending on the seed model, with $\lesssim2~\%$ contribution from gas accretion.}
\label{mergers_vs_accretion}
\end{figure*}
\subsubsection{Mergers vs accretion driven BH growth} 
\label{Mergers vs accretion driven BH growth}
Having discussed the merger rates produced by the different seed models, we shall now look at how much of the BH mass assembly is contributed by mergers versus gas accretion. If we revisit the example subhalos in Figure \ref{BH_wandering_images}, they clearly show that the BH repositioning produces more massive central BHs compared to subgrid DF. In principle, this could be due to a combination of two reasons: 1) more efficient BH mergers, and 2) more efficient accretion as BHs are more likely to spend longer time in regions of dense gas. In Figure \ref{mergers_vs_accretion}, we compare the relative contributions to the BH growth from mergers vs. accretion for a population of BHs at $z=5$, for the different seed models. For repositioning-based simulations, we had already established in \cite{2024MNRAS.531.4311B} that the BH growth is dominated by mergers; we see this in the top panels of Figure \ref{mergers_vs_accretion}. In this paper, we show that this merger-dominated BH growth continues to be true even when we replace the repositioning model by a subgrid DF model~(bottom panels of Figure \ref{mergers_vs_accretion}). Regardless of the seed model, the contribution to the BH mass assembly from gas accretion remains $\lesssim2~\%$. The key difference between the two dynamics treatments would be the amount of BH mass growth that can be sustained by BH mergers. While BH repositioning can grow BH masses at least up to $\sim10^5-10^6~M_{\odot}$ depending on the seed model, the subgrid DF grows BHs only up to masses of $\sim10^4-10^5~M_{\odot}$ due to the reduced merging efficiency. Consequently, we can readily infer that for the BH mass range of $\sim10^3-10^5~M_{\odot}$, the BH mass functions produced by the subgrid-DF model are substantially steeper compared to the repositioning.

The relative dominance of mergers over accretion-driven BH growth of our low mass seeds in high-z halos may arise from the $M_{bh}^2$ scaling of the Bondi accretion formula as well as strong stellar feedback adopted by the TNG model. The latter is found to be a common feature amongst many different simulations~\citep{2016MNRAS.463..529H,2017MNRAS.472L.109A,2021MNRAS.505..172L,2022MNRAS.511..506C}. This essentially identifies a regime where any significant BH mass growth could only be contributed by mergers. The amount of mass growth that occurs in this regime is then determined by the total number of seeds formed as well as the merging efficiency. In Section \ref{Caveats and future work}, we will further discuss how the above results are subject to uncertainties in our modeling of BH accretion and stellar feedback. 

Finally, even though BH growth up to $\sim10^4$–$10^5~M_{\odot}$ is dominated by mergers, the bottom panel of Figure~\ref{mergers_vs_accretion} clearly shows that the contribution from gas accretion increases sharply with BH mass. This suggests that gas accretion is expected to eventually dominate the overall BH mass assembly above a certain threshold mass. To determine this threshold, we plan to run larger-volume simulations in the future, which will allow us to probe the formation and growth of more massive BHs.



\section{Discussion}
\label{discussion}

\subsection{Comparison with subgrid dynamical friction tests of Chen et al. 2022}

Our work is similar in spirit to \cite{2022MNRAS.510..531C} [hereafter C22], who also tested different BH dynamics treatments and merger criteria using the \texttt{Mp-Gadget} code~\citep{2018zndo...1451799F}. There are some key differences between our works as follows: 1) We focus on lower mass $\sim2.2\times10^{3}~M_{\odot}$ seeds forming within dense and metal-poor gas inside $\sim10^6-10^7~M_{\odot}$ halos, whereas C22 produces massive $7.4\times10^{5}~M_{\odot}$ seeds using a halo mass threshold of $10^{10}~M_{\odot}$. 2) Due to differences in spatial resolutions, the C22 simulations trace the BH pairs down to $\sim4~\rm kpc$~(2 times their softening length), whereas we probe down to smaller $\sim 0.2-0.4~\rm kpc$ scales. 3) Finally, C22 applies their own subgrid DF model based on the Chandrashekhar formalism, whereas we use the M23 subgrid DF model.

Despite the above differences, there are several aspects of the C22 results that are consistent with ours. For example, they find that the reduction in the $z\gtrsim5$ merger rates compared to the repositioning model is by a factor of $\sim 6$ when the gravitational bound check is added~(comparing blue vs red solid lines in Figure 11 of C22). This is broadly consistent with our results, particularly for $\seedmassdynamical=64~\seedmass$. Additionally, similar to our results, they also find that when the check for gravitational boundedness is not required for BH-BH mergers, subgrid DF model becomes relatively less consequential to the merger rates. All that being said, the main difference between our results is in the overall predicted merger rates and redshift range of the mergers. Specifically, C22 mergers occur at much later redshifts of $z\lesssim8$ at rates of $\sim2$ per year, where as our mergers start at $z\sim12-20$ at rates of $\sim100-1000$ per year. But this is simply because we resolve much smaller seeds and form them in much smaller~(and abundant) halos at earlier times compared to C22.

Overall, the similarities between our results and C22 are encouraging as they suggest that the different subgrid DF treatments may not produce dramatically different predictions for the merger rates if we adopt similar seed models. However, to properly quantify the modeling uncertainties in the merger rates, it is instructive to perform studies that explore different subgrid DF treatments while keeping all the other physics and numerics the same. We reserve this for future work. 

\subsection{Implications of higher resolution BH dynamics simulations in the literature }
We now put our results in the context of other works in the literature that trace the dynamics of BH pairs to scales smaller than our simulations. We start by noting that our simulations probe the dynamics of BH pairs only down to $\sim0.2-0.4~\rm kpc$, which is the regime wherein dynamical friction is supposed to be the primary contributor to the hardening processes. While we do clearly demonstrate that a large number of pairs can form at those scales, we still have not determined what fraction of these pairs can harden all the way down to the gravitational wave regime. Although we plan to do that in future work, it is instructive to look at several other works that have already studied the dynamics of BH pairs to much smaller scales with resolutions much higher than ours~(albeit with smaller volumes). For example, \cite{2019MNRAS.486..101P} used the \texttt{RAMSES} code to study the evolution BH pairs to a few $\sim10~\rm pc$ scales within idealized and zoom simulations using their subgrid DF model. They generally find that the dynamics of $\lesssim10^5~M_{\odot}$ seeds tend to be erratic at small scales. While their volumes are too small to make statistical predictions for merger rates, one can imagine that the erratic dynamics would certainly have an adverse impact on the merging efficiency. To that end, \cite{2024MNRAS.532.4681P} used high resolution idealized  simulations of merging halos~(hosting BHs), along with \texttt{KETJU} to ``unsoften" the BH interactions to trace the evolution of BH pairs all the way down to GW driven coalescence. They too find that low-mass BHs have difficulty sinking to the halo centers, particularly to within scales well below what our simulations can probe~($\lesssim0.2~\rm kpc$). Additionally, for the BHs that do merge, GW recoils~(not included in our simulations) eject the remnant and make it difficult to undergo future mergers. 

The work of \cite{2021MNRAS.508.1973M} studies the dynamics of seeds in cosmological zoom simulations of massive high-z halos~($\sim10^{12}~M_{\odot}$ at $z\sim5$) using the \texttt{FIRE} galaxy formation model. While we are unable to quantitatively compare our results to theirs~(since we do not plot the exact same quantities), it is still interesting to qualitatively compare their findings as they study the sinking of BHs down to $0.5~\rm kpc$~(similar to our simulations). We also use the same M23 subgrid DF model that was developed and used by the authors of \cite{2021MNRAS.508.1973M}. Notably, while their spatial resolutions are similar to ours, they explicitly resolve the multiphase ISM whereas we use an effective equation of state. They essentially report that seeds below $\sim10^8~M_{\odot}$ find it difficult to sink all the way to the halo centers. 

Our results appear much more optimistic compared to \cite{2021MNRAS.508.1973M} as most of our BHs are located within separations close to the gravitational softening lengths of $\sim0.2-0.4\rm~kpc$~(revisit Figure \ref{spatial_distributions}). Additionally,  we still get a substantial number of merger events at $z>5$~(revisit Figure \ref{merger_rates_gas_based_seeding}). All this could be due to several reasons: First, their simulations probe much more massive $\sim10^{12}~M_{\odot}$ halos at high $z$. They report these halos to be very clumpy, which substantially hampers the sinking of their BHs towards halo centers. Our seeds form and merge in much smaller $\sim10^8~M_{\odot}$ halos, where the halo merger itself can bring the BH pairs to small separations before the BH dynamical friction has to take over. Additionally, the clumpiness in the low-mass high-z halos may be smaller than the most massive high-z halos due to fewer mergers in their recent assembly history; this certainly seems to be the case with the examples visualized in Figure \ref{merger_examples}. With that being said, the clumpiness of our low mass halos could also be underestimated due to the resolution of our simulations as well as the ISM treatment as an  effective equation of state. We shall investigate the impact of resolution and ISM modeling on BH dynamics in future works. Finally, another reason that could contribute to why our BHs seem to be more centered than in \cite{2021MNRAS.508.1973M}, would be differences in our seed models. Specifically, \cite{2021MNRAS.508.1973M} forms seeds out of stars that could be located anywhere in the halo. However, our simulations form seeds out of the densest, metal poor gas cells in halos. As we showed in Section \ref{Seed formation section}, most of our seeds indeed form close to the halo centers already.

While works such as \cite{2019MNRAS.486..101P}, \cite{2021MNRAS.508.1973M}, and \cite{2024MNRAS.532.4681P} have revealed a  number of impediments to BH-BH mergers, several other sub-kpc scale simulations have also revealed processes that facilitate mergers. Specifically, the \texttt{MAGICS} simulation~(Massive black hole Assembly in Galaxies Informed by Cosmological Simulations) suite comprises of high-resolution re-simulations of merging BH pairs extracted from the \texttt{ASTRID} large volume simulations. The  \texttt{MAGICS-I} simulations~\citep{2024OJAp....7E..28C} highlighted how BH pairs could harden rapidly down to $\sim10~\rm pc$ separations if  large scale tidal torques from galaxy mergers could shrink the orbits below 1 kpc within the first $\sim200~\mathrm{Myr}$ after the pair formation in \texttt{ASTRID}. \texttt{MAGICS-II}~\citep{2024arXiv240919914Z} and \texttt{MAGICS-III}~\citep{2024arXiv240919095M} simulations, which added \texttt{KETJU} to trace BH pairs down to coallescence, revealed that the presence of nuclear star clusters (NSCs) is crucial to facilitate rapid sinking and merging of BH pairs, particularly those at $z\gtrsim4$.

Together, all of these small-scale simulations seem to so far suggest that the journey of the BH pairs from $\sim0.2-0.4~\rm kpc$ separations down to the GW regime is challenging, and NSCs may play a key~(perhaps necessary) role in facilitating that. While this is consistent with our assumptions of enhanced dynamical seed masses, we still do not know what fraction of the high-z seed BHs are surrounded by NSCs, and what are the NSC masses; this will be an important direction for future studies. Nevertheless, the abundance of the $\sim0.2-0.4~\rm kpc$ scale BH pairs in our simulations suggests that we can still expect a substantial number of LISA events even if only a small fraction of these pairs end up producing detectable mergers.   

\subsection{Caveats and future work}
\label{Caveats and future work}
There are several caveats to our models that need to be kept in mind while interpreting the results, starting with the uncertainity within subgrid model itself as pointed out by M23. For instance, there is a possibility of double-counting the DF force when the resolution is high enough to naturally resolve \textit{some} of it. M23 suggests an additional sigmoid function that can suppress the subgrid DF to 0 when the BH mass is sufficiently higher than the DM mass~(a regime wherein we expect the DF to be fully resolved). However, idealized tests by \citealt{2024MNRAS.534..957G}~(see Figure 1 of their paper) have shown that even for a BH to DM mass as high as 10000, the DF is not sufficiently resolved for softening lengths as low as $0.14~\mathrm{kpc}$~(which is close to our simulations). Additionally, \cite{2024MNRAS.534..957G} also showed that at fixed BH to DM mass ratio, lower mass halos require higher spatial resolutions to resolve DF. Note that the masses of our seed forming halos~($\sim10^{6}-10^{7}~M_{\odot}$) are much lower than in the tests performed by \citealt{2024MNRAS.534..957G}~($\sim10^{11}-10^{13}~M_{\odot}$). Overall, the results of \citealt{2024MNRAS.534..957G} clearly show that one would require extremely highly spatial and mass resolutions to naturally resolve the DF in the low mass halos wherein the seeds form and evolve in our simulations. Due to these considerations, we decided to simply apply Eq.~(\ref{DF_eqn}) without any additional factor to correct for possible double counting. 

Another potential limitation of the M23 formula is that it only accounts for the DF in a \textit{softened} discrete N-body potential, thereby not including contributions from scattering of background particles with impact parameters below the softening length. Therefore, with this model, we only expect to track the dynamics of BHs down to scales close to the softening lengths~($\sim0.09 \ \rm or \ 0.18~\rm kpc$ depending on the mass resolutions) of the simulations, and not below that. This is distinct from most of the other subgrid DF estimators mentioned earlier, wherein there is an explicit attempt to account for these missing impact parameters. Overall, any subgrid DF estimator will come with its unique set of advantages and disadvantages. Nevertheless, they have all been shown to work reasonably well to capture the small scale BH dynamics, which is a major leap compared to methods that rely on BH repositioning. In this work, we have only explored the M23 estimator. In the future, we shall also use other estimators and study their implications on wandering and merging BH populations predicted by our seed models. 

Next, we do not trace the BH dynamics below the gravitational softening scales. In order to do that, we would need to account for additional hardening processes such as stellar loss cone scattening, viscous gas dissipation from circumbinary gas disks, and GW emission. Therefore, our simulated mergers inevitably underestimate the merger time scales. In addition, we do not include GW induced recoils that can significantly hinder the participation of merger remnants in future generations of mergers. Future work will address all this via standard post-processing approaches~\citep{2017MNRAS.464.3131K,2021MNRAS.501.2531S} to trace the sub-resolution BH binary inspiral, complemented by recently developed models that can do it on-the-fly during the simulation executions~\citep{2024arXiv241007856L,2024arXiv241202374D}. While these approaches will certainly bring us a step closer towards predicting the \textit{actual} rates of GW events, they will inevitably depend on a number of assumptions being made about the unresolved small scale environments of the pairs. In light of these modeling uncertainties, our predicted merger rates would continue to serve as valuable upper limits to the GW event rates that can be used to put constraints on seeding models in the future.

The other set of caveats lie within the assumptions of other aspects of our galaxy formation model that are bound to influence both BH formation and dynamics. For instance, our modeling of star formation and metal enrichment has a substantial influence on the seed formation rates shown in Figure \ref{seed formation}. To that end, we note that many galaxy formation models, including those that inherited the TNG model~\citep{2022MNRAS.511.4005K,2023MNRAS.524.2539P}, under-predict the abundances of the highest redshift~($z\gtrsim12$) galaxies compared to the JWST observations~\citep{2023MNRAS.524.2594K}. Therefore, any changes that we make to the models to account for these JWST observations~(for e.g. higher star formation efficiency or weaker feedback) can have significant consequences to the formation of BH seeds. A weaker stellar feedback could also enhance the accretion rates of low mass seeds at high redshifts and shift the relative importance of merger vs accretion driven BH growth towards the latter. 

Our Bondi-Hoyle prescription also contributes to the difficulty in growing low mass seeds via accretion due to the $M_{bh}^2$ scaling in the accretion rates. Alternative accretion models with weaker scalings between accretion rates and BH mass could substantially increase the accretion rates of low mass seeds~\citep{2017MNRAS.464.2840A,2025arXiv250213241W}. At the same time, magnetic fields can also suppress the actual accretion rates from the Bondi predictions by two orders of magnitude~\citep{2024ApJ...977..200C}. The differences in accretion rates will translate into the differences in the GW event rates originating from BHs of different masses.  Future work will need to address the impact of all these modeling uncertainties, in order to ensure robust constraints on BH seeding using LISA. 

Finally, our small simulation volumes impose several limitations. First, our results are subject to significant cosmic variance. The $[4.5~\mathrm{Mpc}]^3$ boxes were primarily chosen to enable the execution of many runs at a manageable computational cost. Although our final merger rate predictions are based on larger $[9~\mathrm{Mpc}]^3$ boxes, these volumes are still relatively small. Comparisons with larger volume simulations~($[18~\mathrm{Mpc}]^3$ and $[36~\mathrm{Mpc}]^3$) presented in \citealt{2024MNRAS.531.4311B} (see Figure~18) suggest that merger rate estimates may vary by a factor of $\sim2$ due to cosmic variance. Second, as discussed in Section~\ref{Mergers vs accretion driven BH growth}, our small volumes prevent us from studying the formation and growth of more massive BHs, particularly those in the supermassive regime ($\gtrsim10^6~M_{\odot}$). We plan to address these limitations in future work by running larger-volume simulations.     

\section{Summary and Conclusions}
\label{Conclusions}
LISA merger rates are expected to provide the strongest constraints for BH seed models. However, uncertainties in BH dynamics present a major challenge in determining what fraction of seed BHs will participate in merger events. In cosmological simulations, dynamics of low mass seeds is particularly difficult to model due to: 1) inability to resolve the DF, and 2) the~(artificial) numerical heating of BHs due to two-body interactions with massive background particles. In this work, we have studied the dynamics of $\sim10^3~M_{\odot}$ seeds in the high-z~($z\gtrsim5$) Universe using a suite of $[4.5~\mathrm{Mpc}]^3$ and $[9~\mathrm{Mpc}]^3$ cosmological simulations. These boxes add to the \texttt{BRAHMA} suite of simulations that are dedicated to quantify the impact of different BH formation channels on BH populations across the entire cosmic timeline. However, unlike the initial \texttt{BRAHMA} simulations that ``repositioned" the BHs at the local potential minima, our new simulations use a subgrid DF model and trace the dynamics and mergers of BH pairs down to the spatial resolutions~($\sim0.2-0.4\rm~kpc$) of our simulations.   

Our simulations seed $2.2\times10^3~M_{\odot}$ BHs in halos that exceed critical thresholds for dense~\&~metal-poor gas mass~($5-150~\times\seedmass$) and total halo mass~($1000-10000\times \seedmass$). To prevent numerical heating from background particles, our seeds are initialized with enhanced \textit{dynamical seed masses}~($\seedmassdynamical$) that are $3,8,24~\&~64$ times the seed mass; in physical terms, this may represent seeds that are embedded within nuclear star clusters or NSCs. We then perform tests to systematically quantify the impact of removing BH repositioning, and applying subgrid DF under different assumptions for the dynamical seed masses. We also test against simulations that follow the natural dynamics without repositioning or subgrid DF. We study a wide range of properties including the abundances of wandering BHs and spatial and velocity offsets of the BHs with respect to their host halos. Finally, we characterize the impact of different dynamics treatments on the BH merger rates under various criteria for merging BHs based on pair separations as well as gravitational boundedness. Following are our main conclusions:
\begin{itemize}
\item When repositioning is disabled, we find a substantial increase~(by factors of $\sim10$) in the prevalence of BHs wandering at distances of $\sim1$–$10~\rm kpc$ from subhalo centers. This increase occurs regardless of whether subgrid DF is applied. 
\item Subgrid DF is crucial to ensure that a significant portion of the BHs is localized to the subhalo centers. However, the impact of subgrid DF on the BH positions takes time to manifest. At $z\gtrsim15$, not enough time has passed for the subgrid DF to make a substantial impact on the BH kinematics. But by $z\lesssim7$ the subgrid DF ensures that the majority of the BHs are localized to distances of $\sim0.1-0.2~\rm kpc$~(gravitational softening length) from the subhalo centers.

\item The repositioning scheme promptly merges BH pairs at $\sim 2-15~\rm kpc$ scales with associated merger time scales that can be as low as $\lesssim10~\rm Myr$ after the host halo merges. With subgrid DF, we trace the 
inspiraling of the BH pairs down to separations close to two times the gravitational softening lengths~($\sim0.2-0.4~\rm kpc$). Subgrid DF is particularly crucial in ensuring that these inspiralling BH pairs dissipate enough energy to eventually become gravitationally bound and merge. The associated merger time-scales can range from $\sim50-1000~\rm Myr$. In the absence of either subgrid DF or repositioning, the merger time scales are always $\gtrsim200~\rm Myr$. 

\item The delay in BH mergers with subgrid DF leads to a substantial suppression in the merger rates at $z\gtrsim15$, relative to rates with BH repositioning. The suppression is sensitive to the merger redshift and assumed dynamical seed mass. For dynamical seed masses of $3-64$ times the actual seed mass, the $z\gtrsim15$ merger rates are suppressed by factors of $\sim10-100$ respectively. Despite the strong suppression at the highest redshifts, many of the delayed mergers do end up occurring by $z\sim5$. The overall $z\gtrsim5$ merger rates reduce by a factor of $\sim4-10$ compared to predictions using BH repositioning. 

\item For the different seed models, our simulations produce merger rates of $\sim100-1000$ event per year for a dynamical seed mass of $24$ times the actual seed mass. These mergers dominate the earliest stages of the seed BH growth, growing BHs up to $\sim10^4-10^5~M_{\odot}$ by $z\sim5$ with $\lesssim2~\%$ of the BH mass originating from gas accretion.   
\end{itemize}

Overall, our simulations have combined novel gas-based seed models from the \texttt{BRAHMA} simulations with state of the art subgrid DF models to capture the small-scale dynamics of low mass $\sim10^3~M_{\odot}$ seeds. Of course, uncertainties remain in terms of the continued hardening of the BH pairs down to the GW driven regime. Nevertheless, the sheer number of $\sim 0.2-0.4~\rm kpc$ pairs produced in our simulations by $z\sim5$ implies that one can expect a sizable number of LISA events during its life-time even if only a small fraction of these pairs end up actually reaching the GW regime. Future work will study the journey of these BH pairs to smaller scales via processes of stellar loss-cone scattering, viscous dissipation from circumbinary gas disks and GW emission, along with potential ejections of merger remnants due to GW induced recoils.

\section{Acknowledgements}

AKB and PT acknowledge support from NSF-AST 2346977 and the NSF-Simons AI Institute for Cosmic Origins which is supported by the National Science Foundation under Cooperative Agreement 2421782 and the Simons Foundation award MPS-AI-00010515. AKB was also supported in part by grant NSF PHY-2309135 to the Kavli Institute for Theoretical Physics (KITP). Furthermore, AKB acknowledges the organizers of the KITP workshop “Cosmic Origins: The First Billion Years”, during which some of this research was developed. 
LB acknowledges support from NSF award AST-2307171. RW acknowledges funding of a Leibniz Junior Research Group (project number J131/2022).  LH acknowledges support by the Simons Collaboration on ``Learning the Universe''. P.N. acknowledges support from the Gordon and Betty Moore Foundation and the John Templeton Foundation that fund the Black Hole Initiative (BHI) at Harvard University
where she serves as one of the PIs. Finally, the authors acknowledge Research Computing at The University of Virginia and The University of Florida for providing computational resources and technical support that have contributed to the results reported within this publication. URLs: https://rc.virginia.edu~\&~https://www.rc.ufl.edu/

\appendix
\section{Comparing ``default" vs ``higher" resolution boxes for a fixed seeding and dynamics model}
\label{appendix}
\begin{figure*}
\centering
\includegraphics[width= 7cm]{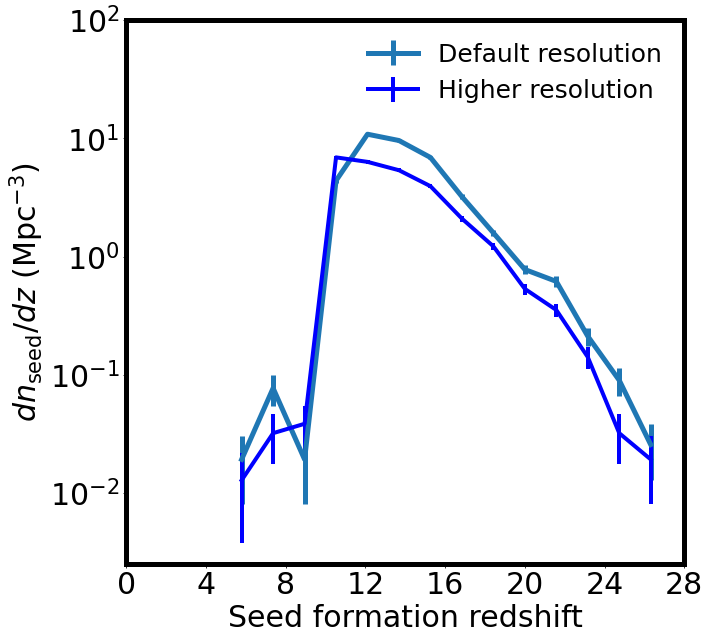} \includegraphics[width= 7cm]{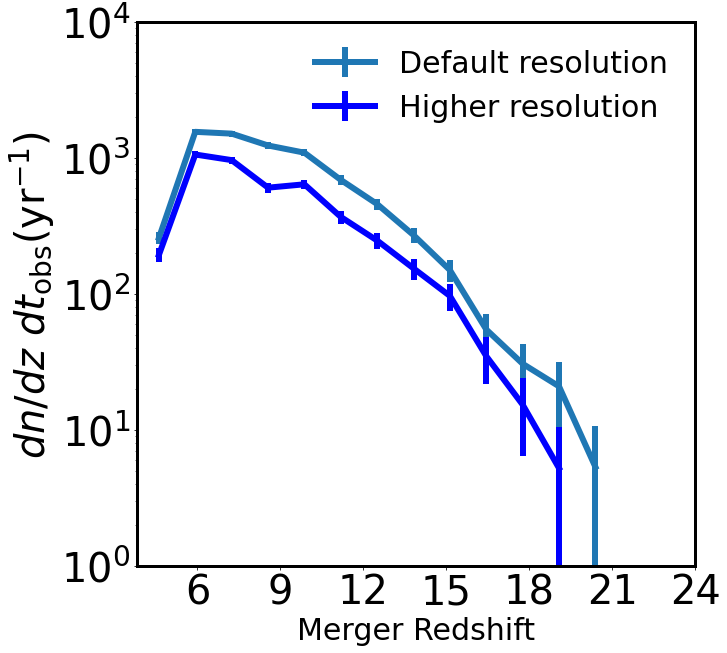}
\caption{Here we compare results for the default resolution and the higher resolution boxes for the \texttt{SM5_FOF1000} seed model and the \texttt{DISCRETE_DF} dynamics model with \texttt{MERGE_SOFTENING_BOUND} and $\seedmassdynamical=64~\seedmass$. The left panels show the seed formation rates and the right panels show the overall merger rates. The higher resolution simulations produce a slightly smaller number of seeds by a factor of $\sim2$, which commensurately also reduces the merger rates.}
\label{resolution_test}
\end{figure*}

In this section, we take a fixed seed model~(\texttt{SM5_FOF1000}) and a fixed dynamics model~(\texttt{DISCRETE_DF} with \texttt{MERGE_SOFTENING_BOUND} and $\seedmassdynamical=64~\seedmass$), and compare the results produced by the default resolution box and the higher resolution box. Recall that the differences between the two boxes are in the DM mass resolution~($2.4\times10^4~M_{\odot}$ vs $3\times10^3~M_{\odot}$) and the softening lengths~($\rm 0.18~kpc$ vs $\rm 0.09~kpc$). The gas mass resolutions are similar amongst the boxes because in the higher resolution box, we de-refined all the gas cells from their initial masses assigned by \texttt{MUSIC}~($\sim f_{\rm baryon}~M_{\rm seed}$) so that they have resolutions similar to the DM particles. This was done to reduce the computational expense as well to avoid the regime wherein the star particle resolution would be comparable to individual massive stars~(our ISM treatment is not appropriate for simulating individual stars). We acknowledge that it would be more ideal to directly generate the gas cells with resolutions similar to DM during the creation of the ICs~(without having to perform a global de-refinement of the gas). However, \texttt{MUSIC} does not have the functionality to do this, and we wanted to use higher resolution version of the same IC realization as the default resolution boxes. To that end, the global de-refinement of the gas cells would very likely induce some ``noise" in the initial power spectrum at the smallest scales. When we look at Figure \ref{resolution_test}~(left panels), we find that the seeding rates in the higher resolution simulations is a factor $\sim2$ lower than the default resolution simulations. This could be due to a difference in the softening length, but also due to the small scale noise possibily induced by the de-refinement. In any case, while this is not negligible, the impact is still relatively small. The merger rates~(right panel of Figure \ref{resolution_test}) are also commensurately reduced only by a factor of $\sim2$ in the higher resolution simulations. Overall, these results encouraged us to proceed forward with using these higher resolution boxes to test the impact of dynamical seed masses throughout the paper.  
\section{Estimating the halo merger time}
\label{halo_merger_estimate}
The accuracy of our halo merger time estimates in Figure \ref{merger_time_scales} is limited by the fact that the halos are output only during the simulation snapshots. Depending on the seed formation times, the BH-BH merger times and the halo merger times, we divide all the merger events into four distinct categories, each with its own method for estimating the halo merger time. These categories are as follows: 
\begin{itemize}
\item In the first category of mergers, there are no snapshots between the secondary BH formation redshift and the BH-BH merger redshift. As a result, the secondary BH is not found in any of the snapshots. In this case, the halo merger could not have occurred earlier than the secondary BH formation redshift~(since halos cannot seed new BHs if they already have one), and also could not have occurred later than the BH-BH merger redshift. Therefore, we approximate the halo merger redshift to be mid-way between the secondary BH formation redshift and the BH-BH merger redshift.

\item In the second category of mergers, there is only one snapshot between the secondary BH formation redshift and the BH-BH merger redshift, and both BHs are found in the same halo at that snapshot. Therefore, the halo merger could not have been earlier than the secondary BH formation redshift, and could not have been later than the snapshot where both BHs are found. The halo merger redshift is then approximated to be at the mid-way between the secondary BH formation redshift and the snapshot redshift.

\item In the third category of mergers, there are at least two snapshots between the secondary BH formation redshift and the BH-BH merger redshift. In this case, the halo merger redshift is approximated to be at the mid-way between two successive snapshots, wherein the earlier snapshot finds the two BHs in different halos and the latter snapshot finds them in the same halo. 

\item Finally, the fourth category comprises of mergers with only one snapshot between the secondary BH formation redshift and the BH-BH merger redshift. But in contrast to the second category, the snapshot does not find the two BHs in the same halo. Therefore, the halo merger could not have happened prior to that snapshot. The halo merger redshift is then approximated to be at the midway between the snapshot and the BH-BH merger redshift. 
\end{itemize}

\bibliography{references}
\end{document}